\documentclass[fleqn,usenatbib]{mnras}

\usepackage{mathptmx}
\usepackage{txfonts}

\usepackage[T1]{fontenc}

\DeclareRobustCommand{\VAN}[3]{#2}
\let\VANthebibliography\thebibliography
\def\thebibliography{\DeclareRobustCommand{\VAN}[3]{##3}\VANthebibliography}

\usepackage{graphicx}	

\newcommand{\MSun}{M$_{\odot}$ }

\newcommand{\mic}{$\mu$m }
\newcommand{\as}{\hbox{$^{\prime\prime}$} }

\title[]{Deep imaging of three accelerating stars using SHARK-NIR and LMIRCam at LBT}

\author[D. Mesa et al.]{
D. Mesa$^{1}$\thanks{E-mail: dino.mesa@inaf.it},
R. Gratton$^{1}$, V. D'Orazi$^{2,1}$, E. Carolo$^{1}$, D. Vassallo$^{1}$,
J. Farinato$^{1}$, L. Marafatto$^{1}$, K. Wagner$^{3}$, \newauthor
J. Hom$^{3}$,
S. Ertel$^{3,4}$, Th. Henning$^{5}$, C. Desgrange$^{5,6}$, D. Barbato$^{1}$,
M. Bergomi$^{1}$, P. Cerpelloni$^{1}$, S. Desidera$^{1}$, \newauthor
S. Di Filippo$^{1}$, D. Doelman$^{7,8}$, 
T.S. Gomes Machado$^{1,9}$, D. Greggio$^{1}$, P. Grenz$^{3}$, M. Kenworthy$^{7}$, F. Laudisio$^{1}$, \newauthor
C. Lazzoni$^{1}$, J. Leisenring$^{3}$,  A. Lorenzetto$^{1}$, 
K.K.R. Santhakumari$^{1}$,
D. Ricci$^{1}$, F. Snik$^{7}$, G. Umbriaco$^{10,1}$, \newauthor
M.C. Vega Pallauta$^{2}$ and V. Viotto$^{1}$
\\
$^{1}$INAF-Osservatorio Astronomico di Padova, Vicolo dell'Osservatorio 5, Padova, Italy, 35122-I\\
$^{2}$Department of Physics, University of Rome Tor Vergata, via della Ricerca Scientifica 1, 00133, Rome, Italy \\
$^{3}$Department of Astronomy and Steward Observatory, The University of Arizona, 933 North Cherry Ave, Tucson, AZ85721, USA \\
$^{4}$Large Binocular Telescope Observatory, The University of Arizona, 933 North Cherry Ave, Tucson, AZ 85721, USA \\
$^{5}$Max Planck Institute for Astronomy, K\"onigstuhl 17, 69117 Heidelberg, Germany\\
$^{6}$Univ. Grenoble Alpes, CNRS, IPAG, F-38000 Grenoble, France \\
$^{7}$Leiden Observatory, Leiden University, P.O. Box 9513, 2300 RA Leiden, the Netherlands \\
$^{8}$SRON Netherlands Institute for Space Research, Niels Bohrweg 4, 2333 CA Leiden, the Netherlands \\
$^{9}$Università degli Studi di Padova - Dipartimento di Fisica e Astronomia, Vicolo dell’ Osservatorio 3, 35122, Padova, Italy \\
$^{10}$Dipartimento di Fisica e Astronomia ”Augusto Righi” - Alma Mater Studiorum Università di Bologna, via Piero Gobetti 93/2 - 40129, Bologna, Italy\\
}

\date{Accepted XXX. Received YYY; in original form ZZZ}

\pubyear{2015}

\begin{document}
\label{firstpage}
\pagerange{\pageref{firstpage}--\pageref{lastpage}}
\maketitle

\begin{abstract}
The combination of detection techniques enhances our ability to identify companions orbiting nearby stars. We employed high-contrast imaging to constrain mass and separation of possible companions responsible for the significant proper motion anomalies of the nearby stars HIP\,11696, HIP\,47110 and HIP\,36277. These targets were observed using the LBT’s high-contrast camera, SHARK-NIR, in H-band using a Gaussian coronagraph, and with the LMIRCam instrument in the L’-band and using a vAPP coronagraph. Both observations were conducted simultaneously. Additionally, constraints at short separations from the host star are derived analyzing the renormalized unit weight error (RUWE) values from the Gaia catalogue. We find that the companion responsible for the anomaly signal of HIP\,11696 is likely positioned at a distance from 2.5 to 28 astronomical units from its host. Its mass is estimated to be between 4 and 16 Jupiter masses, with the greater mass possible only at the upper end of the separation range. Similar limits were obtained for HIP\,47110 where the companion should reside
between 3 and ~30 au with a mass between 3 and 10 MJup. For HIP\,36277, we identified a faint stellar companion at large separation, though it might be substellar depending on the assumed age for the star. Considering the older age, this object accounts for the absolute value of the PMa vector but not for its direction. Additionally, we found a substellar candidate companion at a closer separation that could explain the PMa signal, considering a younger age for the system.

\end{abstract}





\section{Introduction}
\label{intro}

Direct imaging (DI) is the elective technique for detecting massive ($>$1~M$_{\mathrm{Jup}}$) companions at large separation ($>$10~au) from young stars, typically those with ages less than a few hundreds of Myr. The latest generation of high-contrast imagers, e.g., the Gemini Planet imager \citep[GPI;][]{2014PNAS..11112661M}, VLT/SPHERE \citep{2019A&A...631A.155B} and CHARIS \citep{2015SPIE.9605E..1CG} prompted the discovery of an increasing number of such systems. Notable examples include 51\,Eri\,b \citep{2015Sci...350...64M}, HIP\,65426\,b \citep{2017A&A...605L...9C}, PDS\,70\,b \citep{2018A&A...617A..44K,2018A&A...617L...2M}, and PDS\,70\,c \citep{2019NatAs...3..749H}. These discoveries, among others, have underscored the effectiveness of DI in both unveiling new low-mass companions and analyzing their atmospheric properties. \par
However, the paucity of giant planets at the large separations normally explored with DI \citep{2019AJ....158...13N,2021A&A...651A..72V} makes large surveys of young nearby stars very inefficient. These surveys often require the observation of hundreds of targets to detect only a small number of planetary mass objects. Currently, it is believed that the majority of giant planets are located at distances ranging from 1 to 10 au, as evidenced by, e.g., \citet{2018A&A...612L...3M}, who found a peak of 3~au for companions around M stars. However, such separations are generally beyond the capabilities of current high-contrast imaging systems, except for relatively close targets (distances to the Sun approximately less than 50 pc). \par
To address these challenges, targeted observations can be conducted on objects that show indications of having a low-mass companion based on other techniques. For instance, companions initially detected through the radial velocity (RV) technique can subsequently be observed using DI. This approach leverages the strengths of both methods to confirm and further investigate the presence of these companions. Some examples of such synergy are $\beta$\,Pic\,c \citep{2020A&A...642L...2N} and HD\,206893\,c \citep{2023A&A...671L...5H}. \par
Similarly interesting results are obtained by coupling DI data with high-quality astrometric data, thanks in particular to the recent improvement due to the Gaia satellite data releases \citep[e.g.,][]{2023A&A...674A...1G}. Particularly useful to identify promising targets for DI programs is the proper motion anomaly \citep[PMa or acceleration, see e.g.,][]{2018ApJS..239...31B,2019A&A...623A..72K} that is defined as the difference between the short-term and the long-term proper motion measured for a star. The exceptional precision of measurements attained by the Gaia satellite now enables us to gather information about companions in the planetary mass regime. Comparing the Gaia data with those of previous astrometric surveys, e.g., the HIPPARCOS catalogue, different groups compiled catalogues of PMa data \citep{2021ApJS..254...42B,2022A&A...657A...7K}. Recently, this multi-technique approach has allowed us to directly image low-mass companions at short separation from their host stars. One noteworthy case is that of HIP\,99770\,b \citep{2023Sci...380..198C}, a 16~M$_{\mathrm{Jup}}$ companion at a separation of 17~au from its host star. Even more interesting is the case of the companion detected around AF\,Lep \citep{2023A&A...672A..93M,2023A&A...672A..94D,2023ApJ...950L..19F}, a solar mass star around which a planet with a mass of 2-5~M$_{\mathrm{Jup}}$ was imaged at a separation of $\sim$8~au. In addition to these notable instances, other detections of substellar companions were possible thanks to this technique, such as e.g., the brown dwarf HD\,21152\,B \citep{2022MNRAS.513.5588B,2023AJ....165...39F}. Finally, it is important to note that even in the case of non-detection, coupling DI and astrometric data can strongly constrain both separation and mass of the companions causing the PMa signal \citep[e.g.,][]{2022A&A...665A..73M}. \par
Recently, SHARK-NIR \citep{2015IJAsB..14..365F,2022SPIE12185E..22F,2022SPIE12184E..3VM}, a new high-contrast camera, has been installed on the left arm of the Large Binocular Telescope (LBT, Arizona). It can operate in Y, J and H spectral bands, achieving Strehl ratios up to more than 90\% when operating in good weather conditions \citep{2018SPIE10701E..2BC} and is equipped with a set of different coronagraphs, including a classical Gaussian coronagraph, a suite of shaped pupil (SP) coronagraphs and a four-quadrant phase mask. SHARK-NIR is located at the common center-bent Gregorian Focus of the LBT's two apertures together with the Large Binocular Telescope Interferometer \citep[LBTI;][]{2016SPIE.9907E..04H,2020SPIE11446E..07E}
and the SHARK-VIS instrument \citep{2022SPIE12185E..6QP}.  This notably allows for parallel, high-contrast imaging observations with LBT's two 8.4m apertures across the visible light (from B to I spectral bands between 0.4 and 0.9~$\mu$m) with SHARK-VIS, the near infrared (Y to H bands between 0.96 and 1.7~$\mu$m) with SHARK-NIR, the near to mid infrared (J to M bands between 3 and 5~$\mu$m) with LBTI/LMIRCam
\citep{2010SPIE.7735E..3HS,2012SPIE.8446E..4FL} and the N band (8-13~$\mu$m) with LBTI/NOMIC
\citep{2014SPIE.9147E..1OH}.
Considering that a significant portion of high-contrast imagers operating in the near-infrared (NIR) are currently situated in the southern hemisphere, the introduction of SHARK-NIR will open up opportunities to examine regions of the sky that are presently largely uncharted.

Among the science programs selected for the early science of the instrument, we have included the observation of a sample of stars presenting a significant PMa signal from the catalogue described in \citet{2022A&A...657A...7K}. In this paper, we showcase the outcomes from observing the first three targets in our selected sample, illustrating the instrument's ability to significantly constrain the mass and separation of potential companions. This is made possible by its exceptionally deep contrast, which can achieve values as low as 10$^{-6}$ at separations below 1 arcsecond, particularly when observing under optimal conditions. 

The paper is organized as follows. In Section \ref{s:shark} we introduce the high-contrast camera SHARK-NIR. The selection criteria of the sample are described in Section~\ref{s:sample}, alongside a description of the first three targets observed for this program. In Section~\ref{s:obs} we describe our data and the data reduction procedure. In Section~\ref{s:res} we present the results of our observations, which are subsequently discussed in Section~\ref{s:dis}. Finally, in Section~\ref{s:conclusion} we present our conclusions.

\section{SHARK-NIR: the new high-contrast imager at LBT}
\label{s:shark}

SHARK-NIR is a coronagraphic camera operating in the near-infrared spectrum (0.96-1.7~$\mu$m) and equipped with low-resolution spectroscopic capabilities. It is specifically designed to leverage the outstanding performance of the LBT Adaptive Optics system called SOUL \citep{2016SPIE.9909E..3VP,2023arXiv231014447P}. The exceptional angular resolution and contrast are especially crucial for the key science case: the high-contrast imaging observations of 
exoplanets. The instrument features a NIR camera, equipped with a Teledyne H2RG detector. To minimize thermal background interference, the detector is cooled to approximately 76~K using a cryostat. The camera offers a field of view (FOV) roughly 18”x18” and a pixel scale of 14.5 milliarcseconds per pixel. \par
The instrument offers several observing modes to accommodate various astronomical needs. These include:
\begin{itemize}
    \item \textbf{Classical Imaging}: standard imaging mode with no coronagraphs;
    \item \textbf{Coronagraphic Imaging}: this mode exploits different coronagraphic techniques to meet specific requirements in terms of contrast and inner working angle. These include Gaussian, Shaped pupil, and four quadrant phase mask (FQPM) coronagraphic masks;
    \item \textbf{Dual-Band Imaging}: enabled by a Wollaston prism, this mode allows for simultaneous observations in two different bands;
    \item \textbf{Long-Slit Coronagraphic Spectroscopy}: this mode offers spectroscopy with resolutions of R=100 and R=700, covering the entire wavelength range of SHARK-NIR.
\end{itemize}

The simulated performances reveal it can achieve contrasts down to $10^{-6}$ at separations of 300-500 mas \citep{2018SPIE10701E..2BC}. These contrast levels are sufficient to meet the primary scientific objectives, which encompass:
\begin{enumerate}
\item Exoplanet detection and characterization, primarily focusing on extrasolar giant planets located on wide orbits;

\item Circumstellar disks and jets around young stars that are recognized as the site for the planet formation;

\item Solar system objects with the main aim of studying various minor objects within our solar system, including trans-Neptunian and main-belt asteroids;

\item Evolved stars observing the circumstellar environments of stars in the later stages of their life cycles.

\item Active Galactic Nuclei and Quasars: examining the energetic centers of distant galaxies and extremely luminous quasars.

\end{enumerate}

Considering that several science cases expect observations that require performing field de-rotation, the whole instrument is installed on a mechanical bearing. Furthermore, the instrument's design incorporates several subsystems to optimize its performance. In particular, a key addition is the atmospheric dispersion corrector, which facilitates observations at different elevations.
Another significant enhancement is the inclusion of a local deformable mirror (DM, ALPAO DM 97-15), placed in the first pupil plane. This mirror is instrumental in locally minimizing Non-Common Path Aberrations (NCPA) before observation begins. Additionally, when used in its fast tip-tilt mode, the DM serves to correct residual jitter or other movements in the Point Spread Function (PSF) during a scientific exposure. Lastly, this feature allows for the creation of four symmetric spots in the scientific images, which are essential for accurate frame centring (see Section~\ref{s:imaging}).

\section{Sample selection}
\label{s:sample}

We selected our sample from the Hipparcos-Gaia PMa catalogue presented in \citet{2022A&A...657A...7K}. We compiled a list of targets where the signal-to-noise ratio (SNR) of the PMa signal was larger than 3 but less than 20. This specific criterion was established to exclude targets that are likely to have stellar companions.  We selected moreover only stars with a renormalised unit weight error \citep[RUWE;][]{2021A&A...649A...2L} value smaller than 1.4 because a larger value would be indicative of binarity. To image the inner part of the system (less than 10~au), we selected only targets at a distance less than 50~pc. We also chose targets that show definitive signs of youth and have a declination greater than 0$^{\circ}$, making them observable by LBT. Additionally, any stars with obvious indicators of stellar companions (e.g., stars included in binaries catalogs or that in the Gaia catalog have nearby stars with a similar proper motion and parallax), which could account for the PMa signal, were deliberately excluded from our selection. Applying these criteria, we were able to identify several dozen of suitable stars. Among these, during the SHARK-NIR early science program, we observed three targets, namely HIP\,11696,
HIP\,47110 and HIP\,36277 which are the focus of the analysis in this paper and whose main characteristics are detailed in the subsequent sub-sections.

\subsection{HIP\,11696}
\label{s:hip11696}

HIP\,11696 (HD\,15407) is an F5V \citep{2010ApJ...717L..57M} star at a distance of 49.3~pc from the Sun \citep{2023A&A...674A...1G}. The estimated mass is 1.40$\pm$0.07~\MSun \citep{2022A&A...657A...7K}. \citet{2010ApJ...717L..57M} estimated, for this system, an age of 80~Myr, but the same authors found a high probability that this object is part of the AB Doradus Moving Group (ABDMG) as also confirmed more recently by \citet{2018ApJ...860...43G}. The age of this moving group has been estimated to be 125~Myr by \citet{2014AJ....148...70M} using kinematics considerations and $149^{+51}_{-19}$~Myr by \citet{2015MNRAS.454..593B} based on the isochrones. \citet{Gagne2018wd} estimated an age of $117^{+26}_{-13}$~Myr from an analysis of the evolution of the massive white dwarf GD50, that is also a member of this moving group. In this work, we will assume an intermediate age of 137$\pm$12~Myr, following the method recently devised by \citet{2024arXiv240209067G}.

HIP\,11696 has a stellar companion (spectral type K2V and estimated mass of $\sim$0.8~\MSun) at a separation larger than 1000~au \citep{2010ApJ...717L..57M} corresponding to $\sim$20\as, well outside the SHARK-NIR FOV. 
A strong mid-infrared excess was found for this star, hinting towards a recent collision between rocky planetary embryos or planets in the inner part of the system \citep{2010ApJ...717L..57M}. Moreover, a warm (500-600~K) debris disk was detected using {\it Spitzer} data at an estimated separation of 0.6-1.0~au from the central star \citep{2012ApJ...749L..29F}. Such inner disk can help in excluding the presence of massive companions at short separations from the star.\par
Finally, \citet{2022A&A...657A...7K} found a strong PMa signal with a SNR of 4.18, indicative of the presence of a low-mass companion at short separation (few tens of au) from the host star. They estimated for the companion a mass of 6.39~M$_{\mathrm{Jup}}$ if at 3~au from the central star, 4.12~M$_{\mathrm{Jup}}$ at 5~au, 4.93~M$_{\mathrm{Jup}}$ at 10~au and 16.6~M$_{\mathrm{Jup}}$ at 30~au. Because of its large separation and its relatively low mass, we can conclude that this signal is not caused by the stellar companion described above.

\subsection{HIP\,47110}

HIP\,47110 (HD\,82939) is a G5 spectral type star at a distance of 38.7~pc from
the Sun \citep{2023A&A...674A...1G}. Its estimated mass is 0.98$\pm$0.05~\MSun
\citep{2022A&A...657A...7K}. It has been defined as probable member of the
Pleiades Moving group by \citet{2001MNRAS.328...45M} with an estimated age
of 112$\pm$5~Myr \citep{2018ApJ...856...23G}. The star has a stellar companion
at a separation larger than 162\as \citep{2024yCat....102026M} corresponding
to a projected separation larger than 6280~au. \par
\citet{2022A&A...657A...7K} found for this star a PMa signal with a SNR of 3.24
indicating the presence of a companion at low separation with a possible mass
of around 2.5~$M_{\mathrm{Jup}}$ if at a separation between 5 and 10~au and a mass of
11.35~$M_{\mathrm{Jup}}$ if at a separation of 30~au. The large separation of the stellar
companion cited above exclude it as possible cause of the PMa signal.

\subsection{HIP\,36277}

HIP\,36277 is a K3 spectral type star \citep{1985ApJS...59..197B} at a distance
of 46.3~pc from the Sun \citep{2023A&A...674A...1G}. It has an estimated mass
of 0.67~\MSun \citep{2022A&A...657A...7K}. This star was selected in our sample
because it was included in a catalogue of young runaway stars with an age of
$41.2\pm30.7$~Myr by \citet{2011MNRAS.410..190T}. However, a young age seems to
be excluded by the very low abundance of lithium obtained from the analysis of
14 SOPHIE spectra leading to an uncertain age but however older than 1~Gyr. Also
using the PARAM tool \citep{2006A&A...458..609D,2014MNRAS.445.2758R,2017MNRAS.467.1433R} the age of the system appears to be old even if with very large
uncertainties ($5.4\pm5.2$~Gyr). {We will discuss the implications of such age uncertainties in
Section~\ref{s:dis}.\par
The PMa signal obtained by \citet{2022A&A...657A...7K} for this star have a SNR of 3.12 hinting 
toward the presence of a
low mass companion with a mass of 2.3~M$_{\mathrm{Jup}}$ if at a separation of 5~au, of
2.64~M$_{\mathrm{Jup}}$ if at 10~au and of 15.18~M$_{\mathrm{Jup}}$ if at 30~au from the host star.

\section{Observations and data reduction}
\label{s:obs}

\begin{table*}
  \centering
  \caption{Main characteristics of the observations described in this work. For the seeing column we include in parentheses the minimum and the maximum values of the seeing.}
  \label{t:observations}
\begin{tabular}{cccccccc}
\hline
 Target & Obs. date  & Instrument & Coronagraph & Median DIMM seeing & Field rotation & DIT & Total exposure \\
\hline
 HIP\,11696 & 28 Oct. 2023 & SHARK-NIR & Gauss & 1.20\as (0.91\as/ 2.28\as)& $48.6^{\circ}$ & 16.05 s & 3964.35 s\\
 HIP\,11696 & 28 Oct. 2023 & LMIRCam &  vAPP & 1.16\as (0.87\as/ 2.30\as) & $60.9^{\circ}$ & 0.948 s & 5796.22 s \\
 HIP\,47110 & 20 Feb. 2024 & SHARK-NIR & Gauss & N.A.   & $111.76^{\circ}$ & 16.05 s & 4076.85 s \\
 HIP\,47110 & 20 Feb. 2024 & LMIRCam & vAPP &  N.A.     & $138.33^{\circ}$ & 2.53 s  & 6065.38 s  \\
 HIP\,36277 & 21 Feb. 2024 & SHARK-NIR & Gauss & 1.68\as (1.43\as/ 2.15\as) & $33.63^{\circ}$ & 61.53 s & 5660.50 s \\
 HIP\,36277 & 21 Feb. 2024 & LMIRCam & vAPP & 1.82\as (1.16\as/ 2.77\as) & $110.00^{\circ}$ & 5.41 s & 8986.01 s \\
\hline
\end{tabular}
\end{table*}

\begin{figure*}
\centering
\includegraphics[width=0.95\textwidth]{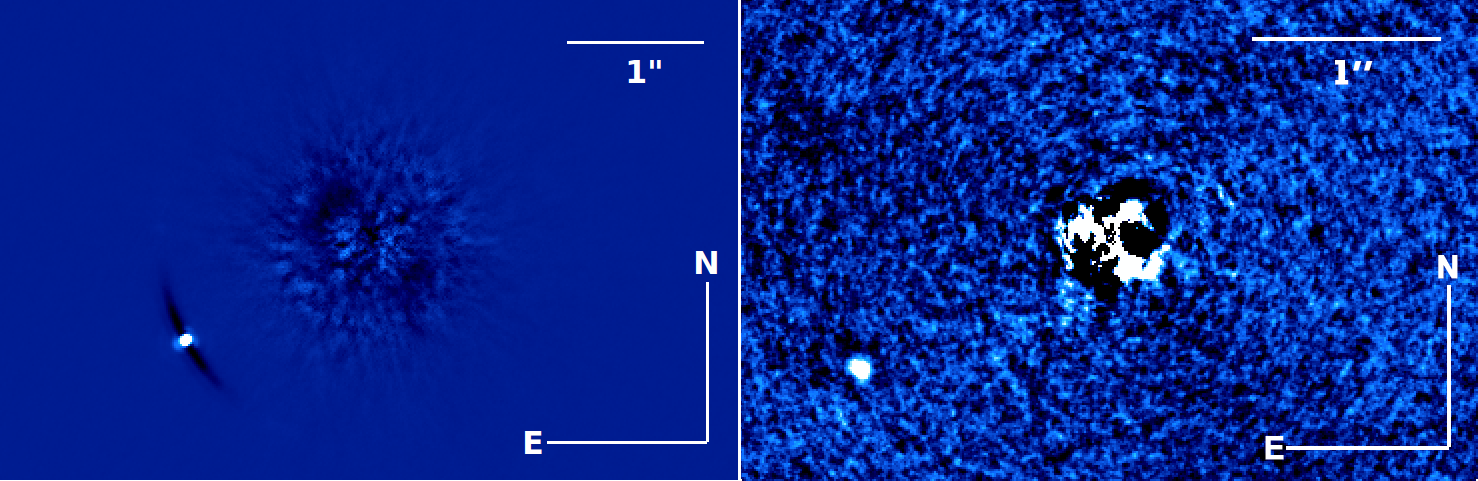}
\caption{{\it Left panel:} Final image obtained for HIP\,11696 using SHARK-NIR   data. This image was obtained by applying a PCA method subtracting 5 principal components. {\it Right panel: } Final image obtained for HIP\,11696 using LMIRCam data. In this case, a PCA method subtracting 10 principal components was applied. In both cases, a bright candidate companion is visible South-East from the star.}
\label{f:hip11696}
\end{figure*}

\subsection{High-contrast imaging data}
\label{s:imaging}

We observed HIP\,11696, HIP\,47110 and HIP\,36277 during the nights of 28
October 2023, 20 February 2024 and 21 February 2024, respectively. All the observations were performed using in parallel the high-contrast imager SHARK-NIR on the left arm and LMIRCam on the right arm of LBT. All the SHARK-NIR observations were performed using the Gaussian coronagraph and the broadband H filter (BB\_H, central wavelength at 1.6~\mic and bandwidth of 0.218~\mic). The simultaneous observation with the LMIRCam instrument were taken in the L' spectral band using the vector-apodizing phase plate coronagraph \citep[vAPP;][]{2021ApOpt..60D..52D}. All the observations were performed in pupil-stabilized mode allowing the rotation of the FOV during the observations to perform angular differential imaging \citep[ADI;][]{2006ApJ...641..556M}. The characteristics of each observation are listed in Table~\ref{t:observations}. For each SHARK-NIR observation, we also took an image with the stellar PSF outside the coronagraph to correctly estimate the contrast for each detected object. This frame was taken by introducing in the optical path an appropriate neutral density filter (ND3) to avoid the saturation of the detector. \par
Because of the non-destructive charge transfer capability of the SHARK-NIR scientific detector, each individual exposure is a ramp, i.e. a set of reads taken at uniform time intervals. The raw ramps were first reduced using a Python pre-processing pipeline specifically developed for the SHARK-NIR detector in collaboration with the University of Arizona. This pre-processing pipeline performs reference pixel correction, bias subtraction, and linearity correction. Finally, the software collapses the ramps into single frames using an up-the-ramp sampling algorithm. This pipeline is important to reduce the readout noise in the final science image and to reduce the possible impact of cosmic rays. \par
The SHARK-NIR data resulting from the pre-processing pipeline described above were then reduced using a scientific pipeline custom designed for the instrument and written in {\tt Python}. First, we subtracted the dark and divided by the flat-field images. For our analysis, we meticulously fine-centered each frame by utilizing the four symmetrical spots created by the SHARK-NIR internal deformable mirror. To this aim, we fitted two lines passing through two opposite spots and considered their intersection as the position of the center of the PSF. This technique ensured precision in aligning the images. Additionally, we conducted a thorough selection process to identify and exclude any frames that were compromised. For example, frames were discarded in instances where, due to residual jitter, the star was not adequately masked by the coronagraph. Finally, we applied a post-processing procedure based on the ADI method and exploited the principal components analysis \citep[PCA;][]{2012ApJ...755L..28S,2012MNRAS.427..948A} algorithm. \par
For what concerns the LMIRCam data, similarly to SHARK-NIR, each exposure is a ramp that in this case is composed of five frames taken at different intermediate exposure times with the first one with a time equal to just 13.7~ms and the last one corresponding to the total exposure time.
As a first step, we subtracted the first image of the ramp data cube from the last one to remove the bias from the science data. The data were taken with two different nodding positions called A and B. We then subtracted from each image a median of the 100 closest images in time from the opposite nod position. This is to subtract thermal background flux from all frames obtained at MIR wavelengths without needing to obtain sky frames. We then aligned the images using cross-correlation and determined the center via rotational centering \citep{Morzinski2015}. We {\it destriped} the images by subtracting the mode from each vertical column of pixels and applied a high pass filter by subtracting a 25-pixel smoothed version of each image from itself. Finally, we applied a post-processing procedure based on PCA similar to what was done for the SHARK-NIR data.


\subsection{Proper motion anomaly data}
\label{s:pmadata}

The PMa data used for this work are obtained from the catalogue by \citet{2022A&A...657A...7K}. Following their approach, we considered all the targets with a PMa SNR larger than 3 as a probable companion-hosting star. Moreover, using the method devised in \citet{2019A&A...623A..72K}, we estimated the mass of the companion that could be responsible of the PMa as a function of the separation from the host star using their Eq.18:
\begin{equation}
    m_2(r) = \frac{\sqrt{2}\epsilon_i}{\varpi r}m_1
\end{equation}
In this formula, $\epsilon_i$ represents the "excess noise" provided in the Gaia catalog, $\varpi$
is the parallax, while $m_1$ and $m_2$ are the masses of the primary and of the companion generating the PMa signal, respectively.
We stress that these mass limits are calculated assuming a circular orbit and they should then be considered as minimum values. It is then important to note that the possible position of the companion causing the PMa signal can be further restricted considering the position angle (PA) of the velocity anomaly vector, as shown in \citet{2022MNRAS.513.5588B}.

\subsection{Radial velocity data}
\label{s:rvdata}

We found 9 public literature and archival radial velocity (RV) observations
for HIP\,36277 collected with the SOPHIE spectrograph and retrieved from the
SOPHIE archive\footnote{\url{atlas.obs-hp.fr/sophie}}. To assess
the RV detection limits, we followed an injection and retrieval scheme similar to the one pursued in, e.g., \citet{barbato2018} and \citet{barbato2023},
injecting synthetic companion signals into the RV time-series and testing for
their significance. To do so, we explored a 40x40 grid of semimajor axes from
0.5 to 50 au and minimum companion mass from 30 M$_\oplus$ to 80~M$_{\mathrm{Jup}}$,
generating for each realization 500 synthetic RV curves with randomly drawn
values of mean longitude $\lambda_0$, eccentricity $e$ and periastron argument
$\omega$. For each of the $8 \cdot 10^5$ synthetic RV time-series we compute
the False Alarm Probability (FAP) value of the injected companion orbital
period, considering it to be detectable if FAP$<10^{-3}$.


\section{Results}
\label{s:res}

\subsection{Detected objects}
\label{s:candidate}
One bright object was imaged in the HIP\,11696 and in the HIP\,36277 data.
In the latter, we also identify another fainter object whose real nature needs
further verification. We did not identify any possible source in the
HIP\,47110 data. We will describe our findings in the following of this
Section. 
\subsubsection{HIP\,11696}
\label{s:HIP11696obj}


\begin{figure}
\centering
\includegraphics[width=\columnwidth]{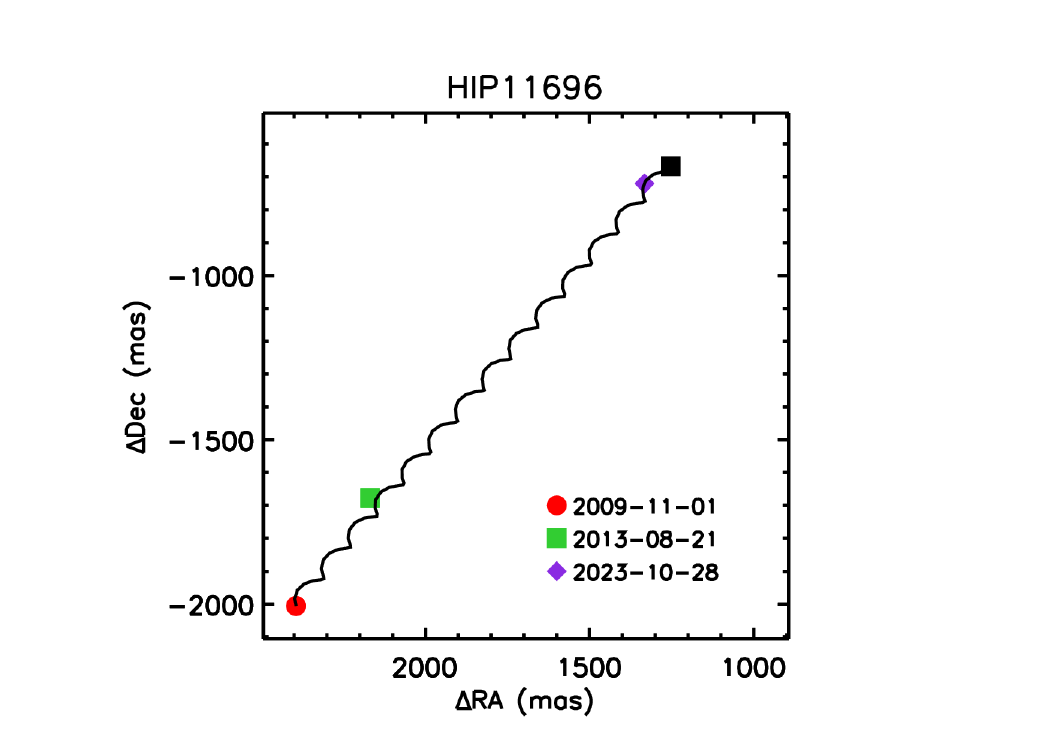}
\caption{Relative astrometric position for the candidate companion detected around HIP\,11696. The red circle represents the relative position of the object at the NIRC2/KeckII observation epoch, while the green square represents the relative position of the candidate companion at the NIRI observation epoch. Finally, the violet diamond represents the relative position at the epoch of the LBT observation. The error bars on the positions at all epochs are not visible in this plot because of their small dimensions. The solid black line represents the expected relative movement for a background object. The black square is the expected position for a background object at the epoch of the SHARK-NIR and LMIRCam observation.}
\label{f:astrohip11696}
\end{figure}

\begin{table*}
  \centering
  \caption{Relative astrometry obtained for the object detected around HIP\,11696.} \label{t:astrohip11696}
\begin{tabular}{ccccc}
\hline
Instrument & Obs. date  & Separation (arcsec) & PA (deg)      & Reference \\
\hline
NIRC2/KeckII&2009-11-01 & 3.213$\pm$0.002     & 129.9$\pm$0.5 &  Galicher et al. 2016 \\
NIRI       & 2013-08-21 & 2.741$\pm$0.015     & 127.7$\pm$0.5 &  Galicher et al. 2016 \\
SHARK-NIR  & 2023-10-28 & 1.514$\pm$0.007     & 118.4$\pm$0.5 & This work \\
LMIRCam    & 2023-10-28 & 1.502$\pm$0.005     & 118.1$\pm$0.5 & This work \\
\hline
\end{tabular}
\end{table*}

The final images obtained for HIP\,11696 are displayed in Figure~\ref{f:hip11696} both for SHARK-NIR (left panel) and LMIRCam (right panel) data. A bright object is visible southeast and at a separation of 1.5\as from the host star in both images. The astrometric values obtained for this source are reported in Table~\ref{t:astrohip11696} together with the astrometric values from previous observations with NIRC2 at the Keck telescope and with NIRI at the Gemini North Telescope relative to the same objects \citep{2016A&A...594A..63G}. In Figure~\ref{f:astrohip11696}, we display the comparison of the relative astrometry of this object at the epoch of the NIRC2 and the NIRI observation (indicated by a red circle and a green square, respectively) and at the epoch of our observation (indicated by a violet diamond). The position of the candidate companion at the epoch of the LBT observation is almost coincident with the expected position of a background object at the same epoch (black square in Figure~\ref{f:astrohip11696}). We can then conclude that this object is actually a background object and that the small discrepancy between the expected position and the measured one is probably due to the intrinsic proper motion of the background star.
We note however that the same object was previously recognized as a background object in 
\citep{2016A&A...594A..63G} so that our data are just a confirmation of their result.

\subsubsection{HIP\,36277}
\label{s:hip36277obj}

\begin{figure*}
\centering
\includegraphics[width=0.95\textwidth]{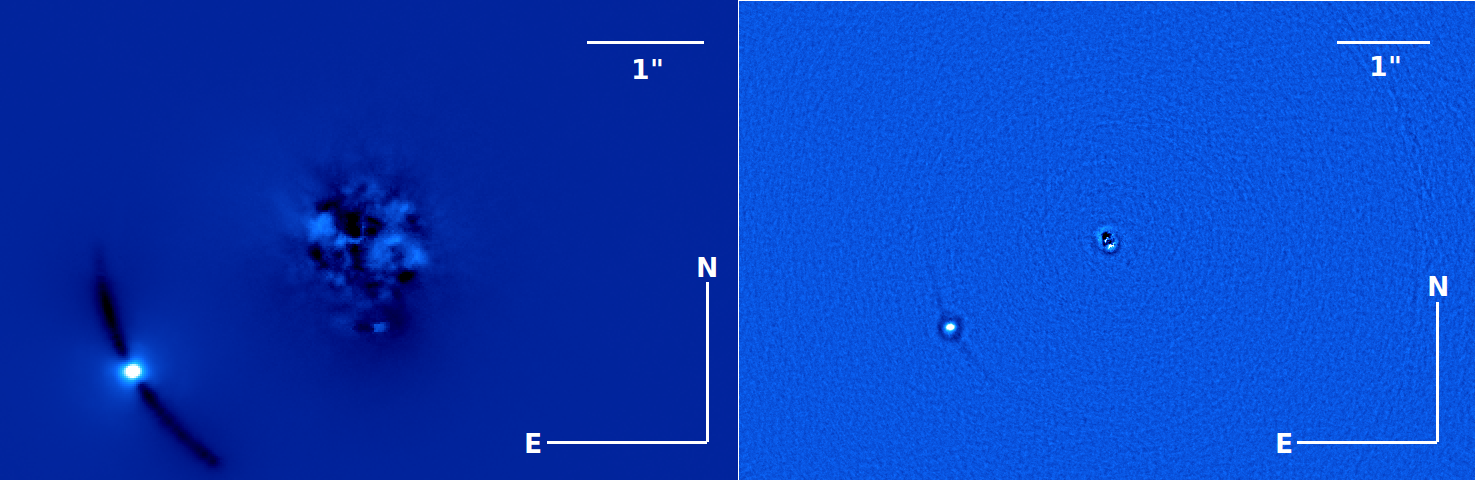}
\caption{{\it Left panel:} Final image obtained for HIP\,36277 using SHARK-NIR   data. This image was obtained by applying a PCA method subtracting 5 principal components. {\it Right panel: } Final image obtained for HIP\,36277 using LMIRCam data. In this case, a PCA method subtracting 10 principal components was applied. In both cases, a bright candidate companion is visible southeast from the star. Furthermore, in the SHARK-NIR image, a possible fainter object is visible just south of the star.}
\label{f:hip36277}
\end{figure*}

\begin{table*}
  \centering
  \caption{Astrometric and photometric values obtained from SHARK-NIR and
    LMIRCam data for the bright external companion of HIP\,36277}
  \label{t:hip36277B}
  \begin{tabular}{cccccc}
    \hline
    Instrument & Sp. band & Separation & PA & Contrast & Absolute magnitude   \\
    \hline
    SHARK-NIR & H & 1.955\as$\pm$0.005\as & $118.9^{\circ}\pm0.1^{\circ}$ & $5.02\pm0.02$~mag & $9.83\pm0.02$~mag \\
    LMIRCam   & L' & 1.927\as$\pm$0.07\as & $118.5^{\circ}\pm0.2^{\circ}$ & $2.87\pm0.05$~mag & $7.53\pm0.05$~mag \\
    \hline
  \end{tabular}
\end{table*}

In Figure~\ref{f:hip36277} we display the final images obtained for HIP\,36277
using both SHARK-NIR (left panel) and LMIRCam (right panel). A bright object is
visible in both images southeast of the star. The same target was also
visible in the Gaia archive with similar parallax and proper motion values
demonstrating physical association \citep{2023A&A...674A...1G}. In
Table~\ref{t:hip36277B} we list the astrometry and the photometry for such
source obtained from our data. Assuming the youngest age of 41~Myr and
using the AMES-COND models \citep{2003IAUS..211..325A} we obtain for this
object a mass of 37.8~M$_{\mathrm{Jup}}$ (with a minimum value of 16.2 and a maximum value
of 73.9~M$_{\mathrm{Jup}}$) while assuming the older age of $\sim$5~Gyr we obtain a mass in
the stellar range just above 0.1~M$_{\odot}$. \par
A faint candidate companion is visible in the SHARK-NIR image with a SNR of 8.66. It is located at a separation of 0.625\as$\pm$0.007\as and a position
angle of $186.9^{\circ}\pm0.2^{\circ}$. At the distance of the system, this
corresponds to a projected separation of around 28.9~au. If real, this object
would have a contrast of $1.43\times10^{-4}$ corresponding to 9.61~mag and 
an absolute H magnitude of 14.42~mag. Assuming an age of 41~Myr and using the AMES-COND models \citep{2003IAUS..211..325A}, this would lead to a mass of 7.60~M$_{\mathrm{Jup}}$ with a minimum value of 3.56~M$_{\mathrm{Jup}}$ and a maximum value of 11.80~M$_{\mathrm{Jup}}$. An older age of
$\sim$5~Gyr would instead lead to a mass in the brown dwarf range with a minimum value of 65.9~M$_{\mathrm{Jup}}$ and a maximum value of 72.1~M$_{\mathrm{Jup}}$. This object was not 
detected in the L' band data obtained with LMIRCam as can be seen in the right panel of Figure~\ref{f:hip36277}. The implications of such non-detection are discussed in Section~\ref{s:sinergy}.

\subsection{Contrast limits}
\label{s:contrast}


\begin{figure*}
\centering
\includegraphics[width=\columnwidth]{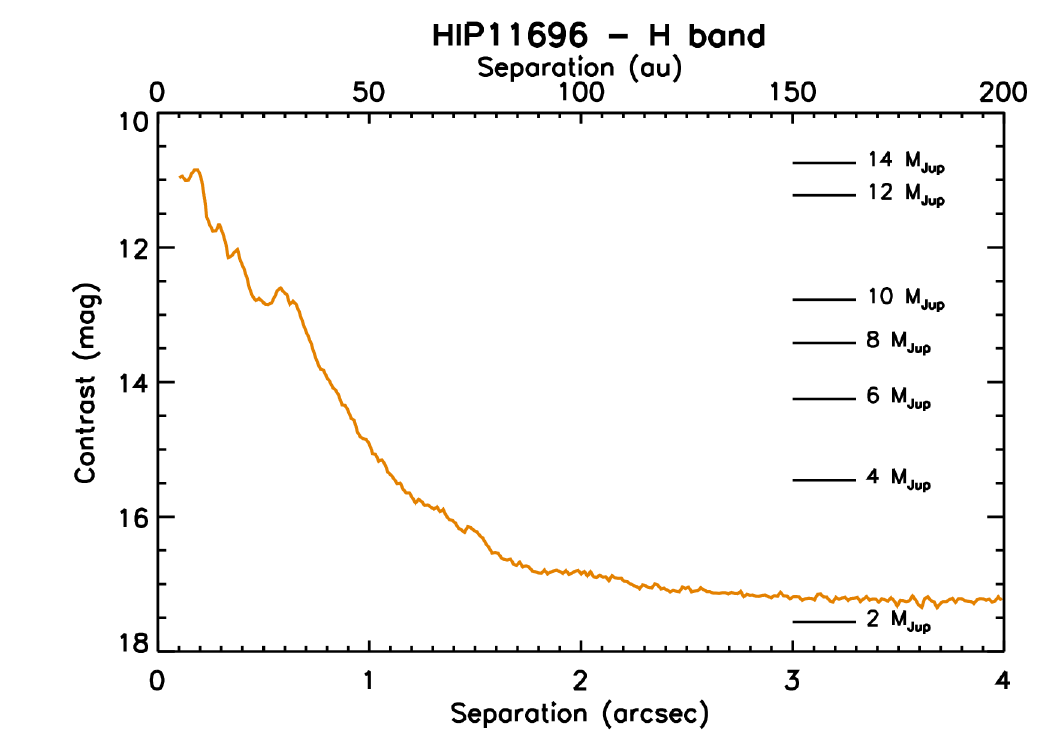}
\includegraphics[width=\columnwidth]{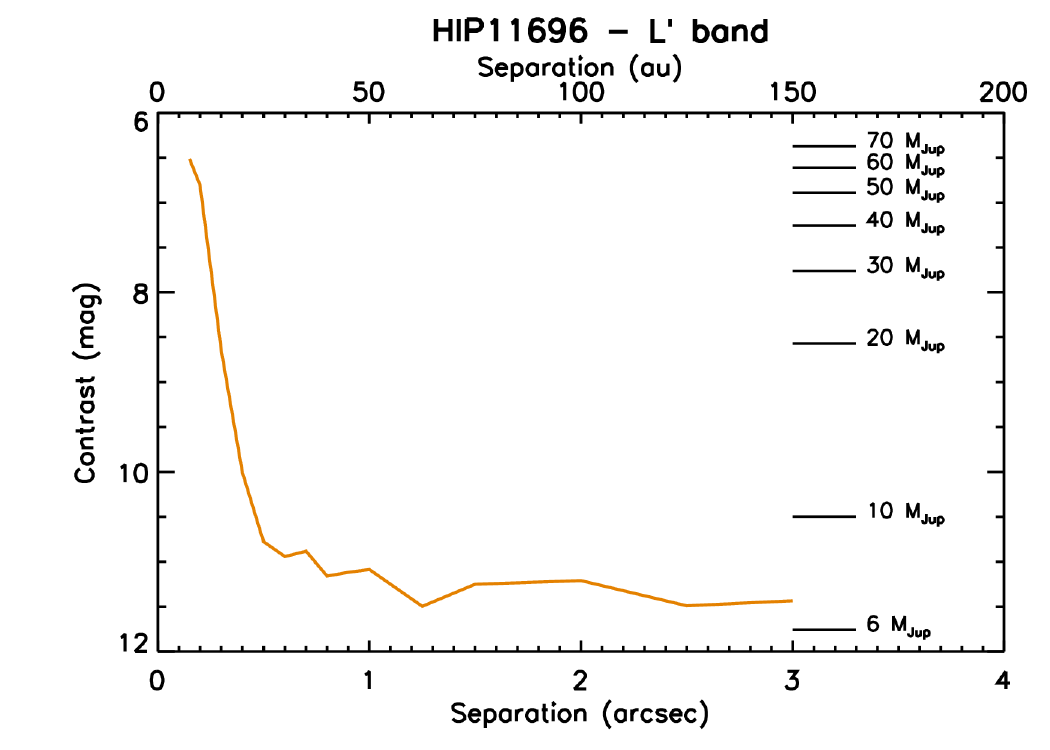}
\includegraphics[width=\columnwidth]{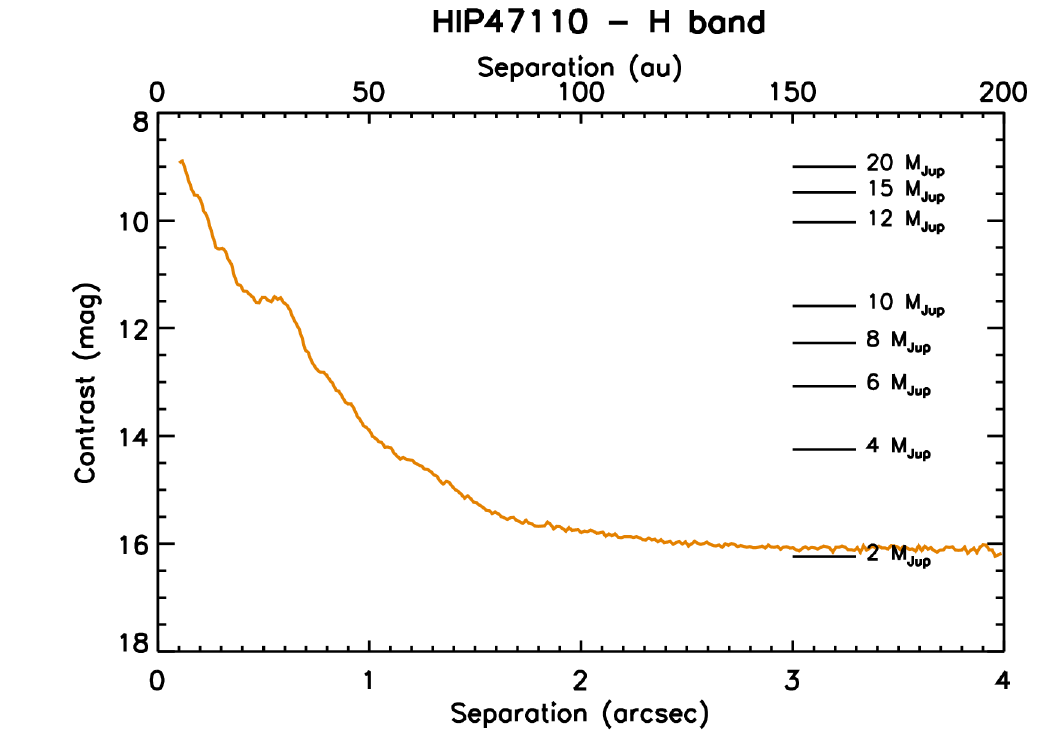}
\includegraphics[width=\columnwidth]{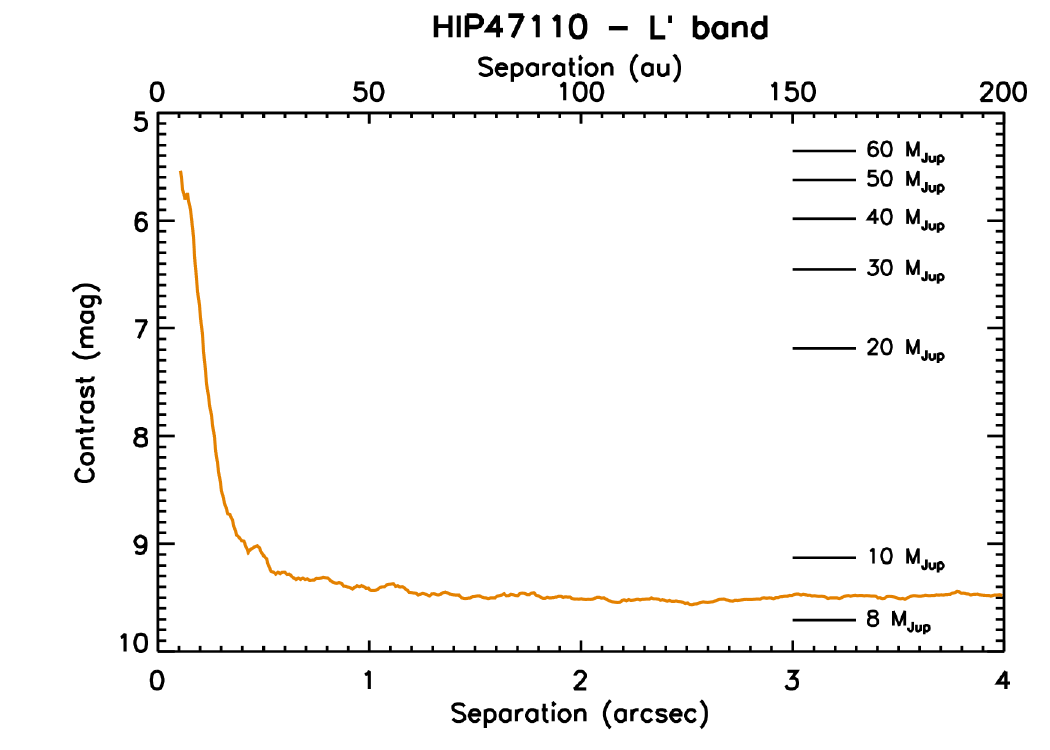}
\includegraphics[width=\columnwidth]{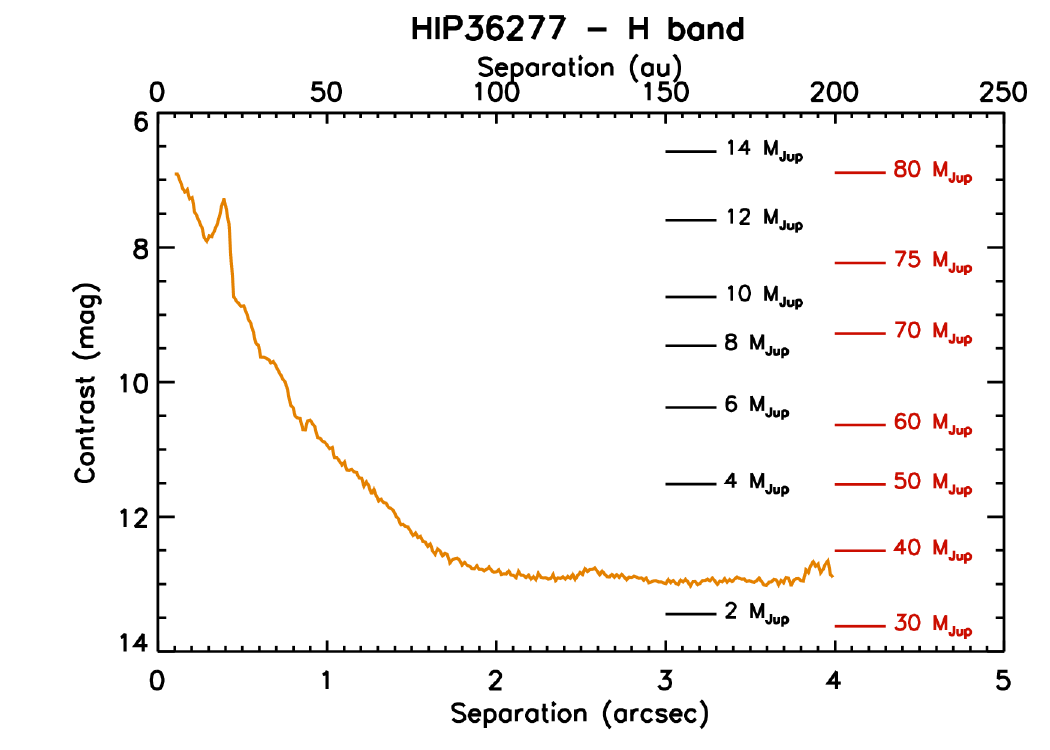}
\includegraphics[width=\columnwidth]{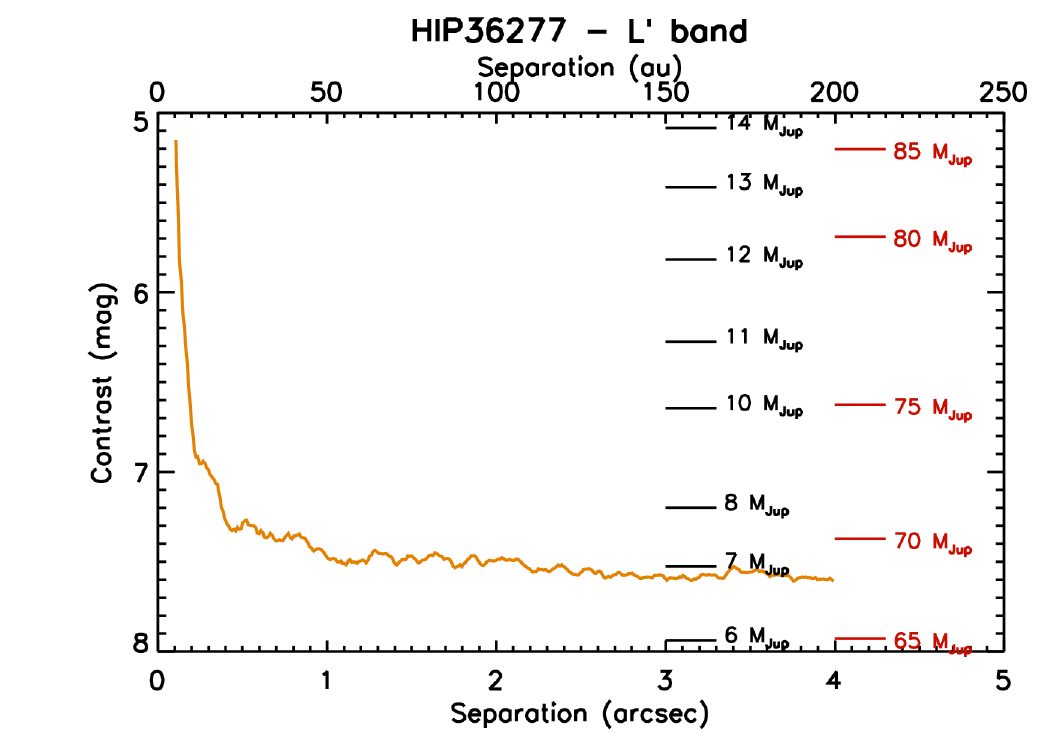}
\caption{Plots of the contrast versus the separation from the hosts star for HIP\,11696(top panels), HIP\,47110 (middle panels) and HIP\,36277 (bottom panels) both for SHARK-NIR (left panels) and LMIRCam (right panels). On the right of each plots we also show the mass limits corresponding to some defined contrasts calculated using the AMES-COND models. See Section~\ref{s:pmacomp} for a discussion on the selection of the atmospheric models. In the case of HIP\,36277 we calculated the mass limits both for the younger age (41~Myr; black lines) and for the older age ($\sim$5~Gyr; red lines).}
\label{f:contrasthip11696}
\end{figure*}



The contrast limits, as a function of the radial distance from the central star, were calculated for both instruments estimating the standard deviation within one-pixel-wide annuli centered on the star. The self-subtraction caused by the post-processing method was evaluated by including simulated objects of known brightness at different separations from the host star. The final contrast was then consequently corrected. Finally, the results were corrected taking into account the effect of the small sample statistics as defined by \citet{2014ApJ...792...97M}. The resulting magnitude contrast limits for HIP\,11696, HIP\,47110 and  HIP\,36277 are shown in the upper, middle and bottom panels of Figure~\ref{f:contrasthip11696}, respectively, both for SHARK-NIR (left panels) and for LMIRCam (right panels). From these plots, we can see that, in the case of
HIP\,11696, with SHARK-NIR we obtained a contrast of around $10^{-5}$ (12.5~mag) at a separation of 0.5\as and of around $10^{-6}$ (15~mag) at a separation of 1\as while a contrast of better than $4\times10^{-7}$ (16~mag) is obtained at separations larger than 1.3\as. On the other hand, LMIRCam reach a contrast of the order of $5\times10^{-5}$ (10.75~mag) at a separation of 0.5\as and of around $3\times10^{-5}$ (11.3~mag) at a separation of 1\as. Slightly worse contrast are obtained for the other two targets mainly because of the worse weather conditions during the observations. Indeed, for HIP\,47110 we obtain with SHARK-NIR a contrast of the order of $2.7\times10^{-5}$ (11.42~mag) at a separation of 0.5\as, a contrast of $\sim2.8\times10^{-6}$ (13.9~mag) at a separation of 1\as and a contrast of $1.3\times10^{-6}$ (14.7~mag) at 1.3\as from the star. Also in this case LMIRCam allowed less deep contrasts of $2.3\times10^{-4}$ (9.10~mag) at a separation of 0.5\as and of $1.7\times10~{-4}$ (9.42~mag) at 1\as from the star. Finally, for HIP\,36277 SHARK-NIR allowed to obtain a contrast of $2.8\times10^{-4}$(8.88~mag) at a separation of 0.5\as, of $4.2\times10^{-5}$ (10.94~mag) at a separation of 1\as and of $1.9\times10^{-5}$ (11.80~mag) at 1.3\as from the star. On the other hand, LMIRCam allows a contrast of $1.2\times10^{-3}$ (7.30~mag) at a separation of 0.5\as and of $1.0\times10^{-3}$ (7.50~mag) at a separation of 1\as. 

\subsection{Synergy between SHARK-NIR and LMIRCam}
\label{s:sinergy}

\begin{figure*}
\centering
\includegraphics[width=\columnwidth]{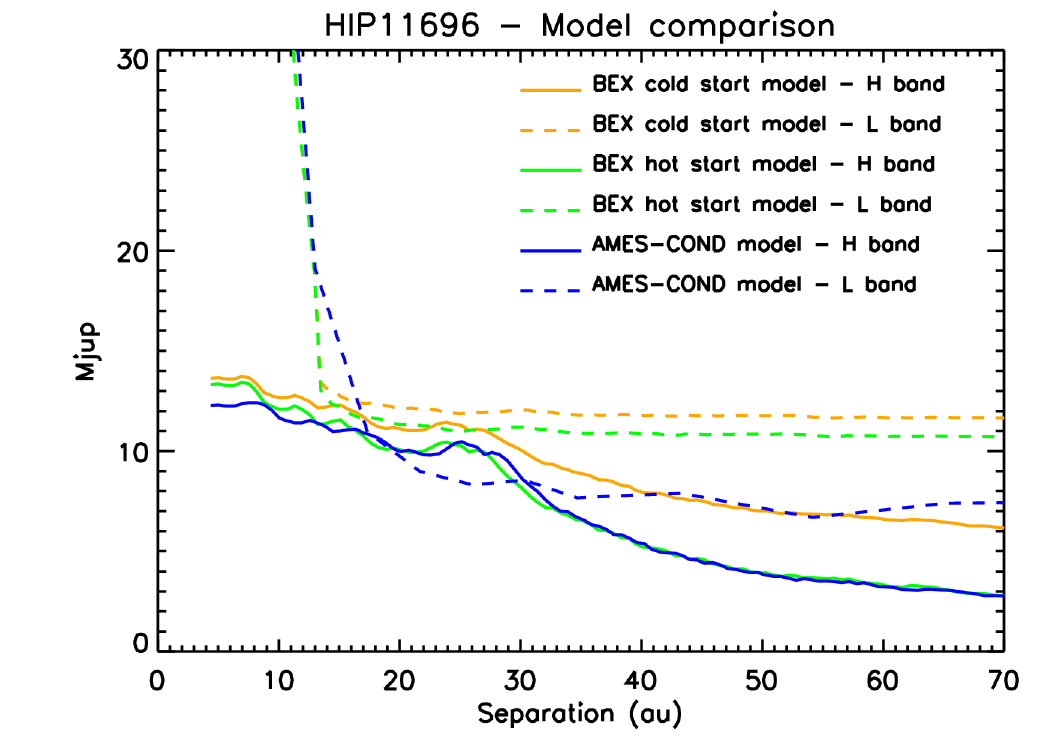}
\includegraphics[width=\columnwidth]{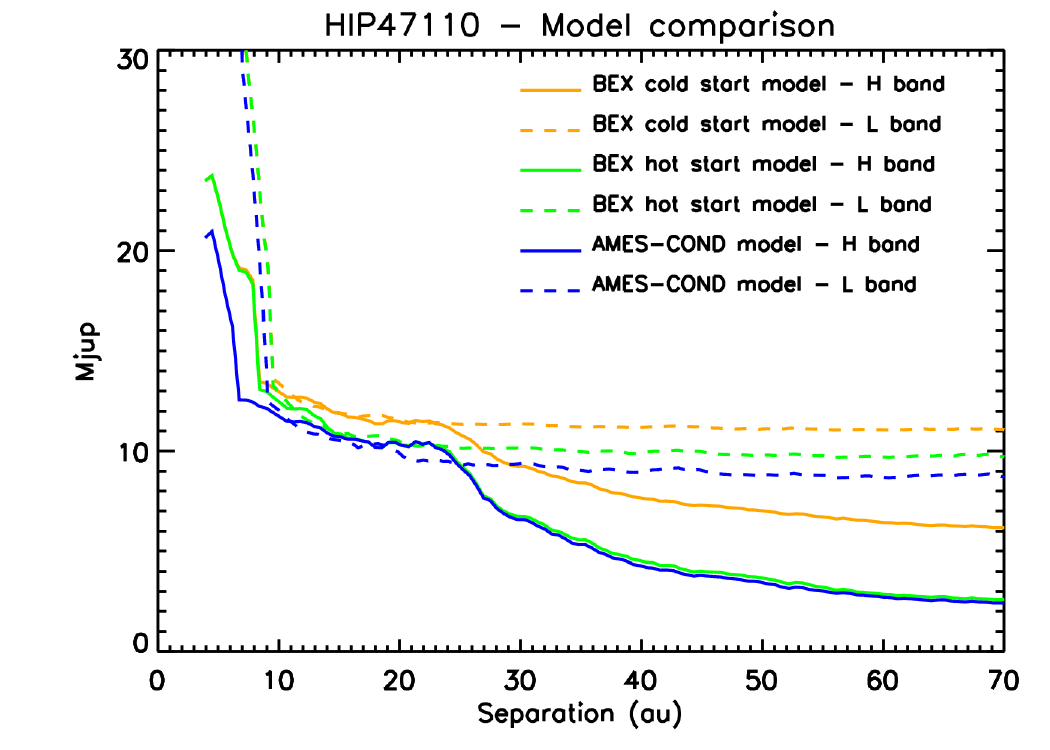}
\includegraphics[width=\columnwidth]{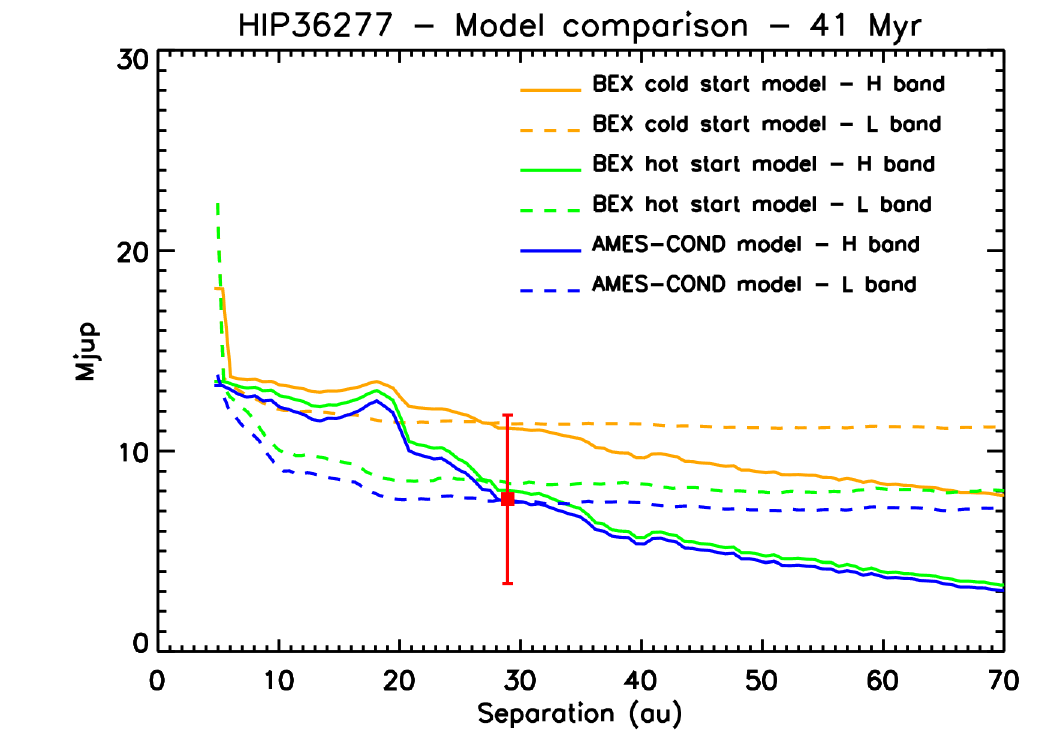}
\includegraphics[width=\columnwidth]{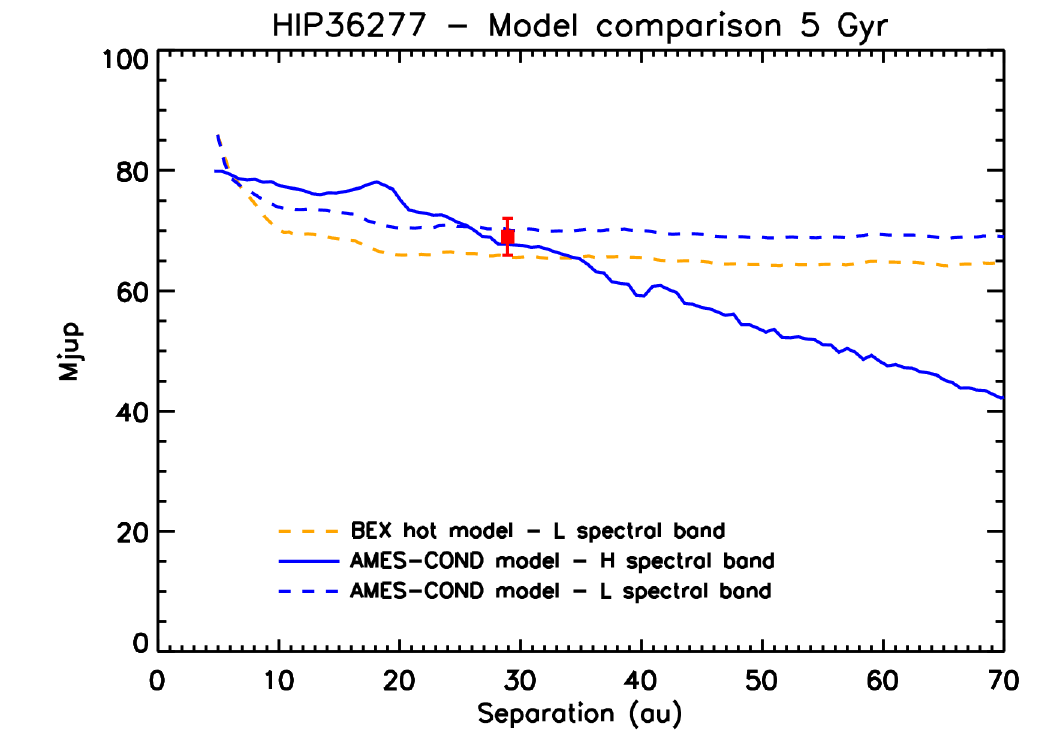}
\caption{Comparison of mass limits for HIP\,11696 (upper left panel), HIP\,47110 (upper right panel) and HIP\,36277 (bottom panels) assuming different atmospheric models both for the H (solid lines) and the L (dashed lines) spectral band. For the case of HIP\,36277 we considered both the younger age of 41~Myr (bottom left panel) and the older age of $\sim$5~Gyr (bottom right panel). In the latter case, as detailed in the text we could use just a lower number of models. For HIP\,36277 we also included the calculated position for the inner faint candidate companion represented by a red square.}
\label{f:modelhip11696}
\end{figure*}

As demonstrated in Section~\ref{s:contrast}, SHARK-NIR offers a deeper contrast than LMIRCam. However, to fully compare the performance of these two instruments, we must also consider their different spectral coverage. The L' band exploited by LMIRCam allows to get lower mass limits with the same contrast obtained by SHARK-NIR in the H spectral band. As a result, the two instruments can provide comparable mass limits that can be favorable to LMIRCam at some specific separations. This can be
seen in Figure~\ref{f:modelhip11696} where better mass limits are obtained using LMIRCam at separations between 20 and 30~au both for HIP\,11696 and HIP\,47110. For HIP\,36277 this is true for an even larger separation range between $\sim$10 and 30~au. This highlights the importance of parallel observations at different wavelength bands aiming  to obtain the best possible mass limits. To demonstrate this,
we have calculated the mass limits for our three targets adopting the stellar ages given in Section~\ref{s:sample} and using the SHARK-NIR and the
LMIRCam data. We used different atmospheric models, namely AMES-COND models, and the cold and hot-start cases from the
Bern EXoplanet cooling curves (BEX) models \citep{2019A&A...624A..20M}. The
results of such procedure are displayed for HIP\,11696 in the upper left panel of 
Figure~\ref{f:modelhip11696}, for HIP\,47110 in the upper right panel of Figure~\ref{f:modelhip11696}
while in the bottom panels of Figure~\ref{f:modelhip11696} 
we display the results for HIP\,36277 for the age of 41.2~Myr (left panel) and 5.3~Gyr (rigth panel),
respectively. For the older age of the latter target, the BEX models are only
available for the L' band. Moreover, as expected the BEX hot-start and cold-start models tends to the same values for such high ages so that their plots are virtually indistinguishable. 
For this reason, we decided to plot just one of them. \par
In general, and as expected, the mass limits obtained using the
cold-start model are worse than those obtained using the hot-start model.
We note that the mass limits obtained with SHARK-NIR are generally better than
those obtained with LMIRCam. 
As demonstrated above, different wavelengths can give better mass limits in different separation
ranges highlighting the importance of observations using different spectral bands.
Moreover, the synergy between the different
wavelength bands could also help in drawing some conclusions about the inner
faint companion detected in the SHARK-NIR image. To this aim we have indicated
with a red square in the bottom panels of Figure~\ref{f:modelhip11696} the expected position for this companion. Its
position is very near to the detection limits or, according to the models,
even lower than such limits both for the H and the L band. Considering only the
AMES-COND models we see that it is at the detection limits in both cases but,
considering the quite large uncertainties on the mass value, it is not
completely surprising that we can detect it only in one wavelength band. As for
the 5~Gyr case, the candidate companion is just above the mass limit in the
H-band case and just below in the L-band case. Then, despite the non-detection
in L' band could be seen as a hint of the non-planetary nature of such an object, the mass limits in H and L spectral bands do not allow us to draw a definitive conclusion about its nature.

\subsection{Comparison of mass limits with PMa results}
\label{s:pmacomp}

\begin{figure}
\centering
\includegraphics[width=\columnwidth]{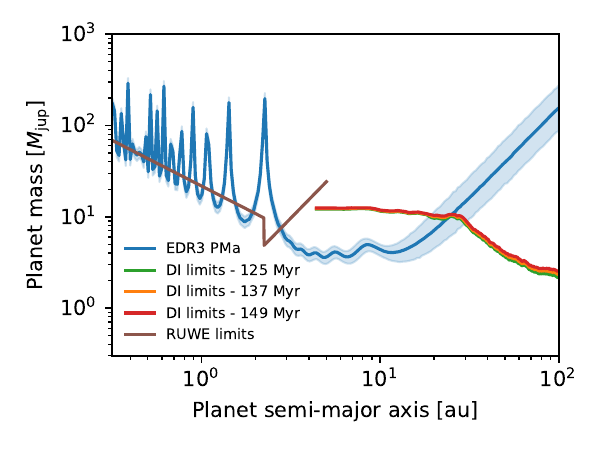}
\caption{Comparison of the plot of mass limits as a function of the separation from the host star to explain the PMa measurement at the Gaia eDR3 epoch (blue lines) for HIP\,11696 with the limits in mass obtained from the SHARK-NIR observations and the AMES-COND atmospheric models (orange solid line). The green and the red solid lines represent the mass limits obtained using the minimum and the maximum ages for the star, respectively. Finally, the brown solid line displays the mass limit at a short separation from the star that can be deduced from considerations of the RUWE value.}
\label{f:pmahip11696}
\end{figure}

\begin{figure}
\centering
\includegraphics[width=\columnwidth]{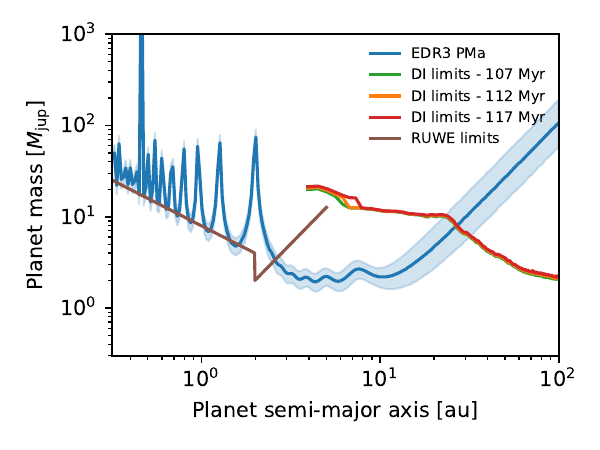}
\caption{Comparison of the plot of mass limits as a function of the separation from the host star to explain the PMa measurement at the Gaia eDR3 epoch (blue lines) for HIP\,47110 with the limits in mass obtained from the SHARK-NIR observations and the AMES-COND atmospheric models (orange solid line). The green and the red solid lines represent the mass limits obtained using the minimum and the maximum ages for the star, respectively. Finally, the brown solid line displays the mass limit at a short separation from the star that can be deduced from considerations of the RUWE value.}
\label{f:pmahip47110}
\end{figure}

\begin{figure}
\centering
\includegraphics[width=\columnwidth]{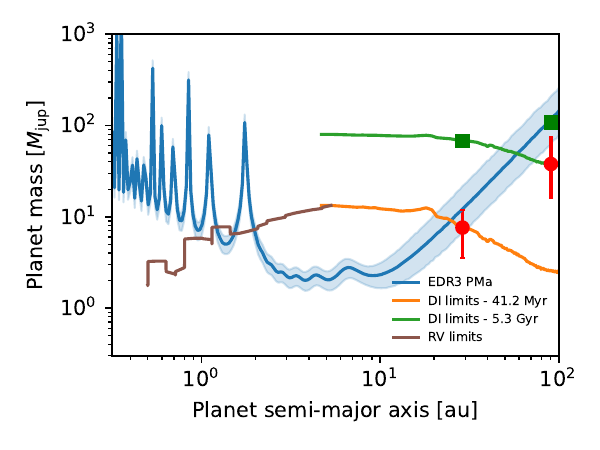}
\caption{Comparison of the plot of mass limits as a function of the separation from the host star to explain the PMa measurement at the Gaia eDR3 epoch (blue lines) for HIP\,36277 with the limits in mass obtained from the SHARK-NIR observations and the AMES-COND atmospheric models. The orange solid line is obtained considering the young age of $\sim$41~Myr while the green solid line represents the limits obtained considering the older age of $\sim$5.3~Gyr. The brown solid line displays the mass limit at a short separation from the star that can be deduced from RV data. The red circles represent the positions of the two candidate companions detected in our images when using the younger age, while the green squares represent the position of the same objects when considering the older age.}
\label{f:pmahip36277}
\end{figure}

We compared the mass limits obtained with the procedure described in Section~\ref{s:sinergy} with the estimates of the mass of the companion responsible for the PMa signal calculated as described in Section~\ref{s:pmadata}. The comparison was done only with the mass limits obtained using the AMES-COND models. This decision was made because the results from these models are comparable to those of the BEX hot-start models. Furthermore, it has been demonstrated \citep{2017A&A...608A..72M} that the hot-start models are in better accordance with the observational data with respect to the cold-start models. A
thorough discussion about the merit of each theoretical model is well outside the goal of this paper but
a further confirmation of this fact derive from the fitting of the data for $\beta$\,Pic moving group 
planetary mass objects with such models done by \citet{2024A&A...684A..69G} (see in particular their Figure~1). This procedure allows us to put constraints on the mass and the separation of the companions responsible for the PMa signal. Further constraints can be defined at small separations from the host star, considering the mass limits obtained using the RUWE value calculated for this star. To this aim, we followed the method devised in \citet{2023A&A...678A..93G}, applying the formulas contained in the Appendix~D of their work.
In the case of HIP\,36277 we can constrain limits at low separation from the star exploiting the RV data presented in Section~\ref{s:rvdata}, so we did not use the RUWE in this case. The results of such procedure are displayed in Figure~\ref{f:pmahip11696}, \ref{f:pmahip47110} and \ref{f:pmahip36277} for
HIP\,11696, HIP\,47110 and HIP\,36277, respectively. For the case of HIP\,36277,
because of the uncertainties on the age of the system, we present both the mass limits obtained assuming the young age of $\sim$41~Myr (orange solid line) and
those obtained assuming the older age of $\sim$5.3~Gyr. We also included the
position in separation and mass for the two candidate companions detected in
our data using a red circle when considering the younger age and a green
square when considering the older age.


\section{Discussion}
\label{s:dis}

\begin{figure}
\centering
\includegraphics[width=\columnwidth]{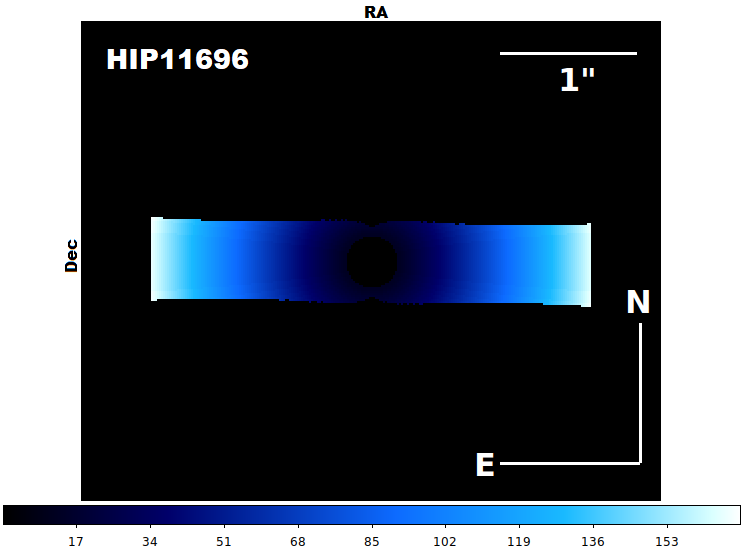}
\includegraphics[width=\columnwidth]{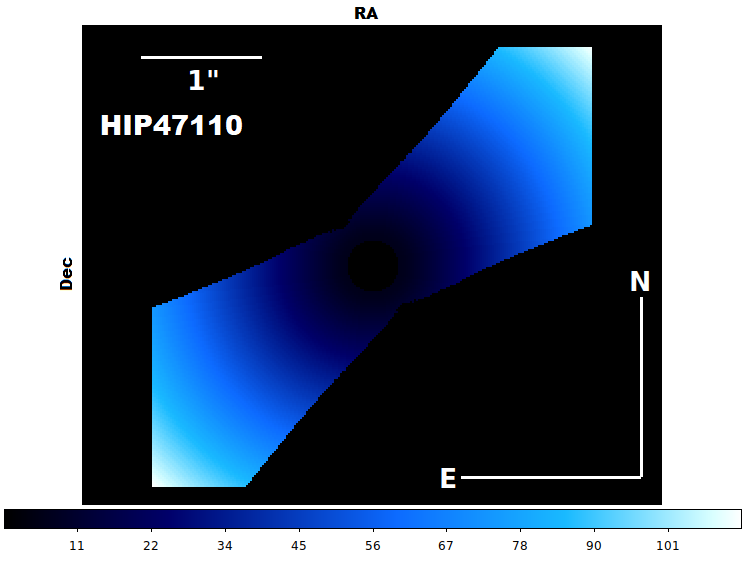}
\includegraphics[width=\columnwidth]{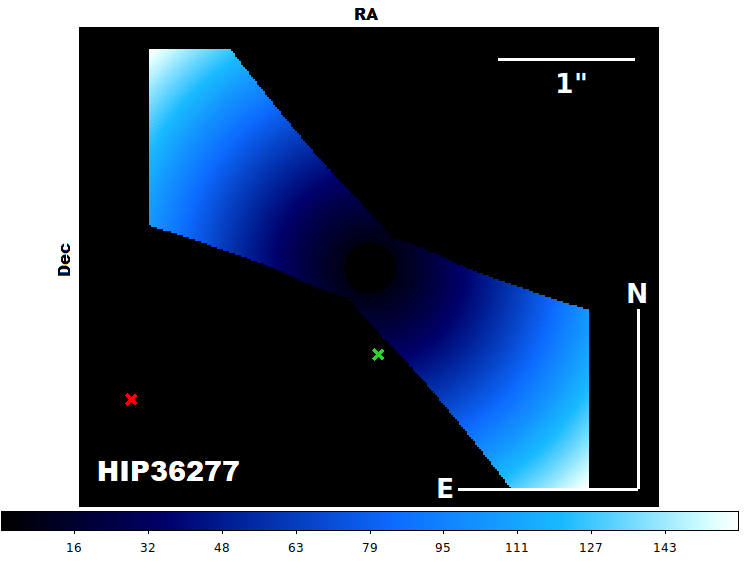}
\caption{2D maps of the FOV around HIP\,11696 (top), HIP\,47110 (middle) and HIP\,36277 (bottom) displaying the sky area compatible with the PMa measured for these stars. The blue-white areas in the images are the regions where the companions causing the PMa signal can reside at the LBT observing epochs. The different colors of these areas represent the masses of the companions generating the PMa signal if they reside in that position. The color scale is indicated in the color bar at the bottom of each image with the masses expressed in M$_{\mathrm{Jup}}$. On the other hand, the black areas in the images represent the regions where the companions generating the PMa signal cannot reside at the moment of the observation. For HIP\,36277, the red and the green crosses are the positions of the two candidates detected in our data. }
\label{f:detmaphip11696}
\end{figure}

The procedure described in the previous Sections allowed us to put tight constraints on the mass and the separation of the companions causing the PMa signal detected for our three targets. For HIP\,11696 and for HIP\,47110, the limits at short separations, obtained taking into account the RUWE values (see Figures~\ref{f:pmahip11696} and \ref{f:pmahip47110}, brown solid line), allow to exclude almost all the possible configurations for host-star separations shorter than 2.5~au and 3~au, respectively. Moreover, the adopted values to calculate those limits are indeed very conservative so we can exclude that the companion generating the PMa signal resides at a small separation from the star. At larger separations, the mass limits obtained through high-contrast imaging allow us to exclude the presence of possible companions at separations larger than 28~au from the host star in the case of HIP\,11696 and at separations larger than 30~au for HIP\,47110. Regarding HIP\,11696 the mass of the companion residing in the small allowed separation range should generally be less than 10~M$_{\mathrm{Jup}}$, with a mean value around 4-5~M$_{\mathrm{Jup}}$ apart for the most external part of the allowed region, at separations above 20~au, where the mass can arrive up to values of around 15-16~M$_{\mathrm{Jup}}$. For HIP\,47110 the masses allowed for the companion generating the PMa signal are even smaller ranging between 2.4 and 10.1~M$_{\mathrm{Jup}}$.
\par
We note that similar observations acquired either with SHARK-NIR or with equivalent high-contrast instruments but in better conditions (like, e.g., with a seeing constantly below 1\as) could allow to further constrain the position and the mass of the companions or even to directly detect them. So these targets are still open to further improvements in their characterization with current instrumentation. \par
The situation is different for the case of HIP\,36277 because of the uncertainties on the real age and the presence of two candidate companions. If we consider the younger age for this system, we could obtain constraints on the mass and separation of companions very similar to those of 
the other targets in our sample (see the orange line in Figure~\ref{f:pmahip47110}). On the 
contrary, if we assume the older age we obtain constraints much less significative as the presence of 
high-mass brown dwarf or low-mass stellar companions would be allowed up to separations of 80~au
(see the green line in Figure~\ref{f:pmahip47110}). The
outer brighter companion, which is confirmed as a bound object thanks to the GAIA data, would have a mass able to explain the PMa signal when considering the
older age proposed for the system while it could not explain it if we consider
the younger age. On the other hand, the PMa signal could be explained by the
inner fainter companion when considering the younger age proposed for the
system. In any case, we stress that this inner companion needs
further confirmation because of its faintness and the low SNR of the
detection. \par
Besides the constraints obtained above using the absolute value of the PMa, we can also take advantage of its vectorial nature and in particular of its position angle, also listed in the catalogue by \citet{2022A&A...657A...7K}, to put limits on the allowed positions for the companion in the FOV of SHARK-NIR. A similar procedure was used for the Finely Optimised REtrieval of Companions of Accelerating STars \citep[FORECAST;][]{2022MNRAS.513.5588B} and it was applied in \citet{2022A&A...665A..73M} and in \citet{2023A&A...672A..93M}.
We then applied the same procedure, adapted for the SHARK-NIR case, for our three targets. The results of the procedure are displayed in Figure~\ref{f:detmaphip11696}. In these images, the region where the companion cannot reside at the moment of the observation is indicated by the dark area. On the other hand, the blue-white colored area represents the region where the companion can be present. The different colors of this area depend on the mass of the companion causing the PMa signal if residing in that position. We would like to stress that, due to the unknown information about the eccentricity and inclination of the companion orbit, this map can only give rough indications about the possible position of the companion. This method can also be used as a finding chart for the companion for future observations with high-contrast imagers. \par
This map is particularly interesting for the case of HIP\,36277. In this case, we have indicated with a red cross the position of the outer bright companion and with a green cross the position of the inner and fainter companion. The position of the outer companion is incompatible with the allowed positions for the companion generating the PMa signal. So, despite there is no doubt about the fact that this object is bound as demonstrated using the Gaia data (see Section~\ref{s:hip36277obj}) and despite its mass could explain the origin of the PMa signal, its position does not match
what is expected, indicating it is not actually responsible for it. On the
other hand, the inner companion is still outside the allowed region but very
near to its edge. We have to stress here that, for the case of low SNR companions, the allowed positions on the detection map should be much more smoothed than what represented in Figure~\ref{f:detmaphip11696}.  Also considering the uncertainties on the map because of the assumptions listed above,
its position cannot exclude it completely as a possible origin of the PMa.
The indications about this system are then quite contradictory and it deserves
further considerations to fully disentangle the possible origin of the
PMa signal as the position of the brighter companion does not match with the expected positions
for the object generating it. This could then be due to the inner fainter companion (if real)
or to a different unseen companion.




\section{Conclusions}
\label{s:conclusion}

In this study, we presented our findings from the high-contrast observations of the nearby and young stars HIP\,11696 and HIP\,47110. Another target, HIP 36277, was initially included due to its presumed youth; however, further analysis has cast doubt on its actual age. All these observations were conducted using the recently installed SHARK-NIR instrument at the Large Binocular Telescope in the H spectral band, as well as with LMIRCam in the L' spectral band. Uniquely, the observations were carried out simultaneously using both the left (SHARK-NIR) and right (LMIRCam) arms of the LBT. Despite less-than-ideal weather conditions during the observations, the deep contrast achieved allowed us to impose strict constraints on the separation and mass of the companions responsible for the high SNR PMa signal detected for HIP\,11696 and HIP\,47110. Our findings suggest that the companion of HIP\,11696 likely has a mass ranging between 4 and 16~M$_{\mathrm{Jup}}$ and is situated at a separation of 2.5 to 28 au from HIP\,11696. For the case of HIP\,47110 the companion probably resides at a separation between 3 and 30~au with a mass ranging between $\sim$2 and 10~M$_{\mathrm{Jup}}$. This analysis corroborates the presence of a planetary-mass companion around each of these two stars, probably located at relatively short separations. Given the distance to these stars, the companion's separation ranges from 0.05\as to 0.6\as. Although the closer separations might be beyond the reach of current high-contrast imaging capabilities, the larger separations are well within the scope of today's technology. This presents exciting opportunities for future refinement in characterizing these systems using existing instruments. \par
As for HIP\,36277, our data allowed the detection of two candidate companions.
While the external and brighter one is confirmed also thanks to a previous
detection with Gaia, the inner one needs further confirmation. New observations
in better weather conditions should allow us to confirm or reject its presence.
The age of this system is disputed and, depending on the adopted estimate, these two sources could be able to explain the absolute value of the PMa signal measured for this
star, but only the inner one has a position consistent with being the source of the observed PMa. This system still presents then some uncertainties that will need further
observations and analysis to be fully solved. \par

Our analysis underscores the importance of these observations, highlighting their value even when no direct detections are made. These studies are crucial in refining and defining potential target samples for future astronomical instruments. Furthermore, they offer valuable constraints that can aid in the strategic planning of observational campaigns, thereby increasing the likelihood of successful detection
\par
Finally, we would like to stress that this work is using for the first time high-contrast imaging data obtained through SHARK-NIR. Despite the non-perfect weather conditions, the instrument was able to deliver deep contrast imaging at very short separations from the host stars allowing to obtain  significant scientific results. This is a good confirmation of the capability of the instrument and it is encouraging for future results exploiting it.  

\section*{Acknowledgements}

We thank Tom Herbst from MPIA-Heidelberg and the LINC-NIRVANA team for 
sharing part of their instrument control SW to operate the motorized axis of 
SHARK-NIR. We also express our appreciation to NASA and Marcia Rieke, the Principal Investigator of JWST/NIRCam, for granting us the opportunity to utilize one of the NIRCam spare detectors as the primary detector for the SHARK-NIR scientific camera. Additionally, we acknowledge the significant contribution of the ALTA Center (alta.arcetri.inaf.it) forecasts, generated using the Astro-Meso-Nh model. Initialization data of the ALTA automatic forecast system is sourced from the
General Circulation Model (HRES) of the European Centre for Medium-Range Weather Forecasts.
The LBT is an international collaboration among institutions in the United States, Italy and Germany. The LBT Corporation partners are: The University of Arizona on behalf of the Arizona university system; Istituto Nazionale di Astrofisica, Italy; LBT Beteiligungsgesellschaft, Germany, representing the Max Planck Society, the Astrophysical Institute Potsdam, and Heidelberg University; The Ohio State University; The Research Corporation, on behalf of The University of Notre Dame, University of Minnesota and the University of Virginia.
This work has made use of data from the European Space Agency (ESA) mission
{\it Gaia} (\url{https://www.cosmos.esa.int/gaia}), processed by
the {\it Gaia} Data Processing and Analysis Consortium (DPAC,
\url{https://www.cosmos.esa.int/web/gaia/dpac/consortium}). Funding for
the DPAC has been provided by national institutions, in particular, the
institutions participating in the {\it Gaia} Multilateral Agreement.
This research has made extensive use of the NASA-ADS, SIMBAD and Vizier databases, operated at CDS, Strasbourg, France. The authors gratefully acknowledge support from the "Programma di Ricerca Fondamentale INAF 2023" of the National Institute of Astrophysics (Large Grant 2023 NextSTEPS and Large Grant 2023 EXODEMO).
V.D. acknowledges the financial contribution from PRIN MUR 2022 (code 2022YP5ACE) funded by the European Union – NextGenerationEU.

\section*{Data Availability}

All the data used for this work are available after a motivated request to the
author of the paper or to the PI of the instrument.



\bibliographystyle{mnras}
\bibliography{pmashark} 

\begin{thebibliography}{}
\makeatletter
\relax
\def\mn@urlcharsother{\let\do\@makeother \do\$\do\&\do\#\do\^\do\_\do\%\do\~}
\def\mn@doi{\begingroup\mn@urlcharsother \@ifnextchar [ {\mn@doi@} {\mn@doi@[]}}
\def\mn@doi@[#1]#2{\def\@tempa{#1}\ifx\@tempa\@empty \href {http://dx.doi.org/#2} {doi:#2}\else \href {http://dx.doi.org/#2} {#1}\fi \endgroup}
\def\mn@eprint#1#2{\mn@eprint@#1:#2::\@nil}
\def\mn@eprint@arXiv#1{\href {http://arxiv.org/abs/#1} {{\tt arXiv:#1}}}
\def\mn@eprint@dblp#1{\href {http://dblp.uni-trier.de/rec/bibtex/#1.xml} {dblp:#1}}
\def\mn@eprint@#1:#2:#3:#4\@nil{\def\@tempa {#1}\def\@tempb {#2}\def\@tempc {#3}\ifx \@tempc \@empty \let \@tempc \@tempb \let \@tempb \@tempa \fi \ifx \@tempb \@empty \def\@tempb {arXiv}\fi \@ifundefined {mn@eprint@\@tempb}{\@tempb:\@tempc}{\expandafter \expandafter \csname mn@eprint@\@tempb\endcsname \expandafter{\@tempc}}}

\bibitem[\protect\citeauthoryear{{Allard}, {Guillot}, {Ludwig}, {Hauschildt}, {Schweitzer}, {Alexander}  \& {Ferguson}}{{Allard} et~al.}{2003}]{2003IAUS..211..325A}
{Allard} F.,  {Guillot} T.,  {Ludwig} H.-G.,  {Hauschildt} P.~H.,  {Schweitzer} A.,  {Alexander} D.~R.,   {Ferguson} J.~W.,  2003, in {Mart{\'\i}n} E.,  ed., ~ Vol. 211, Brown Dwarfs. p.~325

\bibitem[\protect\citeauthoryear{{Amara} \& {Quanz}}{{Amara} \& {Quanz}}{2012}]{2012MNRAS.427..948A}
{Amara} A.,  {Quanz} S.~P.,  2012, \mn@doi [\mnras] {10.1111/j.1365-2966.2012.21918.x}, \href {https://ui.adsabs.harvard.edu/abs/2012MNRAS.427..948A} {427, 948}

\bibitem[\protect\citeauthoryear{{Barbato} et~al.,}{{Barbato} et~al.}{2018}]{barbato2018}
{Barbato} D.,  et~al., 2018, \mn@doi [\aap] {10.1051/0004-6361/201832791}, \href {https://ui.adsabs.harvard.edu/abs/2018A&A...615A.175B} {615, A175}

\bibitem[\protect\citeauthoryear{{Barbato} et~al.,}{{Barbato} et~al.}{2023}]{barbato2023}
{Barbato} D.,  et~al., 2023, \mn@doi [\aap] {10.1051/0004-6361/202345874}, \href {https://ui.adsabs.harvard.edu/abs/2023A&A...674A.114B} {674, A114}

\bibitem[\protect\citeauthoryear{{Bell}, {Mamajek}  \& {Naylor}}{{Bell} et~al.}{2015}]{2015MNRAS.454..593B}
{Bell} C. P.~M.,  {Mamajek} E.~E.,   {Naylor} T.,  2015, \mn@doi [\mnras] {10.1093/mnras/stv1981}, \href {https://ui.adsabs.harvard.edu/abs/2015MNRAS.454..593B} {454, 593}

\bibitem[\protect\citeauthoryear{{Beuzit} et~al.,}{{Beuzit} et~al.}{2019}]{2019A&A...631A.155B}
{Beuzit} J.~L.,  et~al., 2019, \mn@doi [\aap] {10.1051/0004-6361/201935251}, \href {https://ui.adsabs.harvard.edu/abs/2019A&A...631A.155B} {631, A155}

\bibitem[\protect\citeauthoryear{{Bidelman}}{{Bidelman}}{1985}]{1985ApJS...59..197B}
{Bidelman} W.~P.,  1985, \mn@doi [\apjs] {10.1086/191069}, \href {https://ui.adsabs.harvard.edu/abs/1985ApJS...59..197B} {59, 197}

\bibitem[\protect\citeauthoryear{{Bonavita} et~al.,}{{Bonavita} et~al.}{2022}]{2022MNRAS.513.5588B}
{Bonavita} M.,  et~al., 2022, \mn@doi [\mnras] {10.1093/mnras/stac1250}, \href {https://ui.adsabs.harvard.edu/abs/2022MNRAS.513.5588B} {513, 5588}

\bibitem[\protect\citeauthoryear{{Brandt}}{{Brandt}}{2018}]{2018ApJS..239...31B}
{Brandt} T.~D.,  2018, \mn@doi [\apjs] {10.3847/1538-4365/aaec06}, \href {https://ui.adsabs.harvard.edu/abs/2018ApJS..239...31B} {239, 31}

\bibitem[\protect\citeauthoryear{{Brandt}}{{Brandt}}{2021}]{2021ApJS..254...42B}
{Brandt} T.~D.,  2021, \mn@doi [\apjs] {10.3847/1538-4365/abf93c}, \href {https://ui.adsabs.harvard.edu/abs/2021ApJS..254...42B} {254, 42}

\bibitem[\protect\citeauthoryear{{Carolo} et~al.,}{{Carolo} et~al.}{2018}]{2018SPIE10701E..2BC}
{Carolo} E.,  et~al., 2018, in {Creech-Eakman} M.~J.,  {Tuthill} P.~G.,   {M{\'e}rand} A.,  eds,  Society of Photo-Optical Instrumentation Engineers (SPIE) Conference Series Vol. 10701, Optical and Infrared Interferometry and Imaging VI. p. 107012B, \mn@doi{10.1117/12.2312553}

\bibitem[\protect\citeauthoryear{{Chauvin} et~al.,}{{Chauvin} et~al.}{2017}]{2017A&A...605L...9C}
{Chauvin} G.,  et~al., 2017, \mn@doi [\aap] {10.1051/0004-6361/201731152}, \href {https://ui.adsabs.harvard.edu/abs/2017A&A...605L...9C} {605, L9}

\bibitem[\protect\citeauthoryear{{Currie} et~al.,}{{Currie} et~al.}{2023}]{2023Sci...380..198C}
{Currie} T.,  et~al., 2023, \mn@doi [Science] {10.1126/science.abo6192}, \href {https://ui.adsabs.harvard.edu/abs/2023Sci...380..198C} {380, 198}

\bibitem[\protect\citeauthoryear{{De Rosa}, {Nielsen}, {Wahhaj}, {Ruffio}, {Kalas}, {Peck}, {Hirsch}  \& {Roberson}}{{De Rosa} et~al.}{2023}]{2023A&A...672A..94D}
{De Rosa} R.~J.,  {Nielsen} E.~L.,  {Wahhaj} Z.,  {Ruffio} J.-B.,  {Kalas} P.~G.,  {Peck} A.~E.,  {Hirsch} L.~A.,   {Roberson} W.,  2023, \mn@doi [\aap] {10.1051/0004-6361/202345877}, \href {https://ui.adsabs.harvard.edu/abs/2023A&A...672A..94D} {672, A94}

\bibitem[\protect\citeauthoryear{{Doelman} et~al.,}{{Doelman} et~al.}{2021}]{2021ApOpt..60D..52D}
{Doelman} D.~S.,  et~al., 2021, \mn@doi [\ao] {10.1364/AO.422155}, \href {https://ui.adsabs.harvard.edu/abs/2021ApOpt..60D..52D} {60, D52}

\bibitem[\protect\citeauthoryear{{Ertel} et~al.,}{{Ertel} et~al.}{2020}]{2020SPIE11446E..07E}
{Ertel} S.,  et~al., 2020, in {Tuthill} P.~G.,  {M{\'e}rand} A.,   {Sallum} S.,  eds,  Society of Photo-Optical Instrumentation Engineers (SPIE) Conference Series Vol. 11446, Optical and Infrared Interferometry and Imaging VII. p. 1144607, \mn@doi{10.1117/12.2561849}

\bibitem[\protect\citeauthoryear{{Farinato} et~al.,}{{Farinato} et~al.}{2015}]{2015IJAsB..14..365F}
{Farinato} J.,  et~al., 2015, \mn@doi [International Journal of Astrobiology] {10.1017/S1473550414000494}, \href {https://ui.adsabs.harvard.edu/abs/2015IJAsB..14..365F} {14, 365}

\bibitem[\protect\citeauthoryear{{Farinato} et~al.,}{{Farinato} et~al.}{2022}]{2022SPIE12185E..22F}
{Farinato} J.,  et~al., 2022, in {Schreiber} L.,  {Schmidt} D.,   {Vernet} E.,  eds,  Society of Photo-Optical Instrumentation Engineers (SPIE) Conference Series Vol. 12185, Adaptive Optics Systems VIII. p. 1218522, \mn@doi{10.1117/12.2630083}

\bibitem[\protect\citeauthoryear{{Franson} et~al.,}{{Franson} et~al.}{2023a}]{2023AJ....165...39F}
{Franson} K.,  et~al., 2023a, \mn@doi [\aj] {10.3847/1538-3881/aca408}, \href {https://ui.adsabs.harvard.edu/abs/2023AJ....165...39F} {165, 39}

\bibitem[\protect\citeauthoryear{{Franson} et~al.,}{{Franson} et~al.}{2023b}]{2023ApJ...950L..19F}
{Franson} K.,  et~al., 2023b, \mn@doi [\apjl] {10.3847/2041-8213/acd6f6}, \href {https://ui.adsabs.harvard.edu/abs/2023ApJ...950L..19F} {950, L19}

\bibitem[\protect\citeauthoryear{{Fujiwara}, {Onaka}, {Yamashita}, {Ishihara}, {Kataza}, {Fukagawa}, {Takeda}  \& {Murakami}}{{Fujiwara} et~al.}{2012}]{2012ApJ...749L..29F}
{Fujiwara} H.,  {Onaka} T.,  {Yamashita} T.,  {Ishihara} D.,  {Kataza} H.,  {Fukagawa} M.,  {Takeda} Y.,   {Murakami} H.,  2012, \mn@doi [\apjl] {10.1088/2041-8205/749/2/L29}, \href {https://ui.adsabs.harvard.edu/abs/2012ApJ...749L..29F} {749, L29}

\bibitem[\protect\citeauthoryear{{Gagn{\'e}} et~al.,}{{Gagn{\'e}} et~al.}{2018a}]{2018ApJ...856...23G}
{Gagn{\'e}} J.,  et~al., 2018a, \mn@doi [\apj] {10.3847/1538-4357/aaae09}, \href {https://ui.adsabs.harvard.edu/abs/2018ApJ...856...23G} {856, 23}

\bibitem[\protect\citeauthoryear{{Gagn{\'e}}, {Roy-Loubier}, {Faherty}, {Doyon}  \& {Malo}}{{Gagn{\'e}} et~al.}{2018b}]{2018ApJ...860...43G}
{Gagn{\'e}} J.,  {Roy-Loubier} O.,  {Faherty} J.~K.,  {Doyon} R.,   {Malo} L.,  2018b, \mn@doi [\apj] {10.3847/1538-4357/aac2b8}, \href {https://ui.adsabs.harvard.edu/abs/2018ApJ...860...43G} {860, 43}

\bibitem[\protect\citeauthoryear{{Gagn{\'e}}, {Fontaine}, {Simon}  \& {Faherty}}{{Gagn{\'e}} et~al.}{2018c}]{Gagne2018wd}
{Gagn{\'e}} J.,  {Fontaine} G.,  {Simon} A.,   {Faherty} J.~K.,  2018c, \mn@doi [\apjl] {10.3847/2041-8213/aacdff}, \href {https://ui.adsabs.harvard.edu/abs/2018ApJ...861L..13G} {861, L13}

\bibitem[\protect\citeauthoryear{{Gaia Collaboration} et~al.,}{{Gaia Collaboration} et~al.}{2023}]{2023A&A...674A...1G}
{Gaia Collaboration} et~al., 2023, \mn@doi [\aap] {10.1051/0004-6361/202243940}, \href {https://ui.adsabs.harvard.edu/abs/2023A&A...674A...1G} {674, A1}

\bibitem[\protect\citeauthoryear{{Galicher} et~al.,}{{Galicher} et~al.}{2016}]{2016A&A...594A..63G}
{Galicher} R.,  et~al., 2016, \mn@doi [\aap] {10.1051/0004-6361/201527828}, \href {https://ui.adsabs.harvard.edu/abs/2016A&A...594A..63G} {594, A63}

\bibitem[\protect\citeauthoryear{{Gratton} et~al.,}{{Gratton} et~al.}{2023}]{2023A&A...678A..93G}
{Gratton} R.,  et~al., 2023, \mn@doi [\aap] {10.1051/0004-6361/202346806}, \href {https://ui.adsabs.harvard.edu/abs/2023A&A...678A..93G} {678, A93}

\bibitem[\protect\citeauthoryear{{Gratton} et~al.,}{{Gratton} et~al.}{2024a}]{2024arXiv240209067G}
{Gratton} R.,  et~al., 2024a, \mn@doi [arXiv e-prints] {10.48550/arXiv.2402.09067}, \href {https://ui.adsabs.harvard.edu/abs/2024arXiv240209067G} {p. arXiv:2402.09067}

\bibitem[\protect\citeauthoryear{{Gratton} et~al.,}{{Gratton} et~al.}{2024b}]{2024A&A...684A..69G}
{Gratton} R.,  et~al., 2024b, \mn@doi [\aap] {10.1051/0004-6361/202348012}, \href {https://ui.adsabs.harvard.edu/abs/2024A&A...684A..69G} {684, A69}

\bibitem[\protect\citeauthoryear{{Groff} et~al.,}{{Groff} et~al.}{2015}]{2015SPIE.9605E..1CG}
{Groff} T.~D.,  et~al., 2015, in {Shaklan} S.,  ed.,  Society of Photo-Optical Instrumentation Engineers (SPIE) Conference Series Vol. 9605, Techniques and Instrumentation for Detection of Exoplanets VII. p. 96051C, \mn@doi{10.1117/12.2188465}

\bibitem[\protect\citeauthoryear{{Haffert}, {Bohn}, {de Boer}, {Snellen}, {Brinchmann}, {Girard}, {Keller}  \& {Bacon}}{{Haffert} et~al.}{2019}]{2019NatAs...3..749H}
{Haffert} S.~Y.,  {Bohn} A.~J.,  {de Boer} J.,  {Snellen} I.~A.~G.,  {Brinchmann} J.,  {Girard} J.~H.,  {Keller} C.~U.,   {Bacon} R.,  2019, \mn@doi [Nature Astronomy] {10.1038/s41550-019-0780-5}, \href {https://ui.adsabs.harvard.edu/abs/2019NatAs...3..749H} {3, 749}

\bibitem[\protect\citeauthoryear{{Hinkley} et~al.,}{{Hinkley} et~al.}{2023}]{2023A&A...671L...5H}
{Hinkley} S.,  et~al., 2023, \mn@doi [\aap] {10.1051/0004-6361/202244727}, \href {https://ui.adsabs.harvard.edu/abs/2023A&A...671L...5H} {671, L5}

\bibitem[\protect\citeauthoryear{{Hinz} et~al.,}{{Hinz} et~al.}{2016}]{2016SPIE.9907E..04H}
{Hinz} P.~M.,  et~al., 2016, in {Malbet} F.,  {Creech-Eakman} M.~J.,   {Tuthill} P.~G.,  eds,  Society of Photo-Optical Instrumentation Engineers (SPIE) Conference Series Vol. 9907, Optical and Infrared Interferometry and Imaging V. p. 990704, \mn@doi{10.1117/12.2233795}

\bibitem[\protect\citeauthoryear{{Hoffmann}, {Hinz}, {Defr{\`e}re}, {Leisenring}, {Skemer}, {Arbo}, {Montoya}  \& {Mennesson}}{{Hoffmann} et~al.}{2014}]{2014SPIE.9147E..1OH}
{Hoffmann} W.~F.,  {Hinz} P.~M.,  {Defr{\`e}re} D.,  {Leisenring} J.~M.,  {Skemer} A.~J.,  {Arbo} P.~A.,  {Montoya} M.,   {Mennesson} B.,  2014, in {Ramsay} S.~K.,  {McLean} I.~S.,   {Takami} H.,  eds,  Society of Photo-Optical Instrumentation Engineers (SPIE) Conference Series Vol. 9147, Ground-based and Airborne Instrumentation for Astronomy V. p. 91471O, \mn@doi{10.1117/12.2057252}

\bibitem[\protect\citeauthoryear{{Keppler} et~al.,}{{Keppler} et~al.}{2018}]{2018A&A...617A..44K}
{Keppler} M.,  et~al., 2018, \mn@doi [\aap] {10.1051/0004-6361/201832957}, \href {https://ui.adsabs.harvard.edu/abs/2018A&A...617A..44K} {617, A44}

\bibitem[\protect\citeauthoryear{{Kervella}, {Arenou}, {Mignard}  \& {Th{\'e}venin}}{{Kervella} et~al.}{2019}]{2019A&A...623A..72K}
{Kervella} P.,  {Arenou} F.,  {Mignard} F.,   {Th{\'e}venin} F.,  2019, \mn@doi [\aap] {10.1051/0004-6361/201834371}, \href {https://ui.adsabs.harvard.edu/abs/2019A&A...623A..72K} {623, A72}

\bibitem[\protect\citeauthoryear{{Kervella}, {Arenou}  \& {Th{\'e}venin}}{{Kervella} et~al.}{2022}]{2022A&A...657A...7K}
{Kervella} P.,  {Arenou} F.,   {Th{\'e}venin} F.,  2022, \mn@doi [\aap] {10.1051/0004-6361/202142146}, \href {https://ui.adsabs.harvard.edu/abs/2022A&A...657A...7K} {657, A7}

\bibitem[\protect\citeauthoryear{{Leisenring} et~al.,}{{Leisenring} et~al.}{2012}]{2012SPIE.8446E..4FL}
{Leisenring} J.~M.,  et~al., 2012, in {McLean} I.~S.,  {Ramsay} S.~K.,   {Takami} H.,  eds,  Society of Photo-Optical Instrumentation Engineers (SPIE) Conference Series Vol. 8446, Ground-based and Airborne Instrumentation for Astronomy IV. p. 84464F, \mn@doi{10.1117/12.924814}

\bibitem[\protect\citeauthoryear{{Lindegren} et~al.,}{{Lindegren} et~al.}{2021}]{2021A&A...649A...2L}
{Lindegren} L.,  et~al., 2021, \mn@doi [\aap] {10.1051/0004-6361/202039709}, \href {https://ui.adsabs.harvard.edu/abs/2021A&A...649A...2L} {649, A2}

\bibitem[\protect\citeauthoryear{{Macintosh} et~al.,}{{Macintosh} et~al.}{2014}]{2014PNAS..11112661M}
{Macintosh} B.,  et~al., 2014, \mn@doi [Proceedings of the National Academy of Science] {10.1073/pnas.1304215111}, \href {https://ui.adsabs.harvard.edu/abs/2014PNAS..11112661M} {111, 12661}

\bibitem[\protect\citeauthoryear{{Macintosh} et~al.,}{{Macintosh} et~al.}{2015}]{2015Sci...350...64M}
{Macintosh} B.,  et~al., 2015, \mn@doi [Science] {10.1126/science.aac5891}, \href {https://ui.adsabs.harvard.edu/abs/2015Sci...350...64M} {350, 64}

\bibitem[\protect\citeauthoryear{{Marafatto} et~al.,}{{Marafatto} et~al.}{2022}]{2022SPIE12184E..3VM}
{Marafatto} L.,  et~al., 2022, in {Evans} C.~J.,  {Bryant} J.~J.,   {Motohara} K.,  eds,  Society of Photo-Optical Instrumentation Engineers (SPIE) Conference Series Vol. 12184, Ground-based and Airborne Instrumentation for Astronomy IX. p. 121843V, \mn@doi{10.1117/12.2629494}

\bibitem[\protect\citeauthoryear{{Marleau}, {Coleman}, {Leleu}  \& {Mordasini}}{{Marleau} et~al.}{2019}]{2019A&A...624A..20M}
{Marleau} G.-D.,  {Coleman} G. A.~L.,  {Leleu} A.,   {Mordasini} C.,  2019, \mn@doi [\aap] {10.1051/0004-6361/201833597}, \href {https://ui.adsabs.harvard.edu/abs/2019A&A...624A..20M} {624, A20}

\bibitem[\protect\citeauthoryear{{Marois}, {Lafreni{\`e}re}, {Doyon}, {Macintosh}  \& {Nadeau}}{{Marois} et~al.}{2006}]{2006ApJ...641..556M}
{Marois} C.,  {Lafreni{\`e}re} D.,  {Doyon} R.,  {Macintosh} B.,   {Nadeau} D.,  2006, \mn@doi [\apj] {10.1086/500401}, \href {https://ui.adsabs.harvard.edu/abs/2006ApJ...641..556M} {641, 556}

\bibitem[\protect\citeauthoryear{{Mason}, {Wycoff}, {Hartkopf}, {Douglass}  \& {Worley}}{{Mason} et~al.}{2024}]{2024yCat....102026M}
{Mason} B.~D.,  {Wycoff} G.~L.,  {Hartkopf} W.~I.,  {Douglass} G.~G.,   {Worley} C.~E.,  2024, {VizieR Online Data Catalog: The Washington Visual Double Star Catalog (Mason+ 2001-2020)}, VizieR On-line Data Catalog: B/wds. Originally published in: 2001AJ....122.3466M

\bibitem[\protect\citeauthoryear{{Mawet} et~al.,}{{Mawet} et~al.}{2014}]{2014ApJ...792...97M}
{Mawet} D.,  et~al., 2014, \mn@doi [\apj] {10.1088/0004-637X/792/2/97}, \href {https://ui.adsabs.harvard.edu/abs/2014ApJ...792...97M} {792, 97}

\bibitem[\protect\citeauthoryear{{McCarthy} \& {Wilhelm}}{{McCarthy} \& {Wilhelm}}{2014}]{2014AJ....148...70M}
{McCarthy} K.,  {Wilhelm} R.~J.,  2014, \mn@doi [\aj] {10.1088/0004-6256/148/4/70}, \href {https://ui.adsabs.harvard.edu/abs/2014AJ....148...70M} {148, 70}

\bibitem[\protect\citeauthoryear{{Melis}, {Zuckerman}, {Rhee}  \& {Song}}{{Melis} et~al.}{2010}]{2010ApJ...717L..57M}
{Melis} C.,  {Zuckerman} B.,  {Rhee} J.~H.,   {Song} I.,  2010, \mn@doi [\apjl] {10.1088/2041-8205/717/1/L57}, \href {https://ui.adsabs.harvard.edu/abs/2010ApJ...717L..57M} {717, L57}

\bibitem[\protect\citeauthoryear{{Mesa} et~al.,}{{Mesa} et~al.}{2022}]{2022A&A...665A..73M}
{Mesa} D.,  et~al., 2022, \mn@doi [\aap] {10.1051/0004-6361/202244033}, \href {https://ui.adsabs.harvard.edu/abs/2022A&A...665A..73M} {665, A73}

\bibitem[\protect\citeauthoryear{{Mesa} et~al.,}{{Mesa} et~al.}{2023}]{2023A&A...672A..93M}
{Mesa} D.,  et~al., 2023, \mn@doi [\aap] {10.1051/0004-6361/202345865}, \href {https://ui.adsabs.harvard.edu/abs/2023A&A...672A..93M} {672, A93}

\bibitem[\protect\citeauthoryear{{Meyer}, {Amara}, {Reggiani}  \& {Quanz}}{{Meyer} et~al.}{2018}]{2018A&A...612L...3M}
{Meyer} M.~R.,  {Amara} A.,  {Reggiani} M.,   {Quanz} S.~P.,  2018, \mn@doi [\aap] {10.1051/0004-6361/201731313}, \href {https://ui.adsabs.harvard.edu/abs/2018A&A...612L...3M} {612, L3}

\bibitem[\protect\citeauthoryear{{Montes}, {L{\'o}pez-Santiago}, {G{\'a}lvez}, {Fern{\'a}ndez-Figueroa}, {De Castro}  \& {Cornide}}{{Montes} et~al.}{2001}]{2001MNRAS.328...45M}
{Montes} D.,  {L{\'o}pez-Santiago} J.,  {G{\'a}lvez} M.~C.,  {Fern{\'a}ndez-Figueroa} M.~J.,  {De Castro} E.,   {Cornide} M.,  2001, \mn@doi [\mnras] {10.1046/j.1365-8711.2001.04781.x}, \href {https://ui.adsabs.harvard.edu/abs/2001MNRAS.328...45M} {328, 45}

\bibitem[\protect\citeauthoryear{{Mordasini}, {Marleau}  \& {Molli{\`e}re}}{{Mordasini} et~al.}{2017}]{2017A&A...608A..72M}
{Mordasini} C.,  {Marleau} G.~D.,   {Molli{\`e}re} P.,  2017, \mn@doi [\aap] {10.1051/0004-6361/201630077}, \href {https://ui.adsabs.harvard.edu/abs/2017A&A...608A..72M} {608, A72}

\bibitem[\protect\citeauthoryear{{Morzinski} et~al.,}{{Morzinski} et~al.}{2015}]{Morzinski2015}
{Morzinski} K.~M.,  et~al., 2015, \mn@doi [\apj] {10.1088/0004-637X/815/2/108}, \href {https://ui.adsabs.harvard.edu/abs/2015ApJ...815..108M} {815, 108}

\bibitem[\protect\citeauthoryear{{M{\"u}ller} et~al.,}{{M{\"u}ller} et~al.}{2018}]{2018A&A...617L...2M}
{M{\"u}ller} A.,  et~al., 2018, \mn@doi [\aap] {10.1051/0004-6361/201833584}, \href {https://ui.adsabs.harvard.edu/abs/2018A&A...617L...2M} {617, L2}

\bibitem[\protect\citeauthoryear{{Nielsen} et~al.,}{{Nielsen} et~al.}{2019}]{2019AJ....158...13N}
{Nielsen} E.~L.,  et~al., 2019, \mn@doi [\aj] {10.3847/1538-3881/ab16e9}, \href {https://ui.adsabs.harvard.edu/abs/2019AJ....158...13N} {158, 13}

\bibitem[\protect\citeauthoryear{{Nowak} et~al.,}{{Nowak} et~al.}{2020}]{2020A&A...642L...2N}
{Nowak} M.,  et~al., 2020, \mn@doi [\aap] {10.1051/0004-6361/202039039}, \href {https://ui.adsabs.harvard.edu/abs/2020A&A...642L...2N} {642, L2}

\bibitem[\protect\citeauthoryear{{Pedichini} et~al.,}{{Pedichini} et~al.}{2022}]{2022SPIE12185E..6QP}
{Pedichini} F.,  et~al., 2022, in {Schreiber} L.,  {Schmidt} D.,   {Vernet} E.,  eds,  Society of Photo-Optical Instrumentation Engineers (SPIE) Conference Series Vol. 12185, Adaptive Optics Systems VIII. p. 121856Q, \mn@doi{10.1117/12.2629244}

\bibitem[\protect\citeauthoryear{{Pinna} et~al.,}{{Pinna} et~al.}{2016}]{2016SPIE.9909E..3VP}
{Pinna} E.,  et~al., 2016, in {Marchetti} E.,  {Close} L.~M.,   {V{\'e}ran} J.-P.,  eds,  Society of Photo-Optical Instrumentation Engineers (SPIE) Conference Series Vol. 9909, Adaptive Optics Systems V. p. 99093V, \mn@doi{10.1117/12.2234444}

\bibitem[\protect\citeauthoryear{{Pinna} et~al.,}{{Pinna} et~al.}{2023}]{2023arXiv231014447P}
{Pinna} E.,  et~al., 2023, \mn@doi [arXiv e-prints] {10.48550/arXiv.2310.14447}, \href {https://ui.adsabs.harvard.edu/abs/2023arXiv231014447P} {p. arXiv:2310.14447}

\bibitem[\protect\citeauthoryear{{Rodrigues} et~al.,}{{Rodrigues} et~al.}{2014}]{2014MNRAS.445.2758R}
{Rodrigues} T.~S.,  et~al., 2014, \mn@doi [\mnras] {10.1093/mnras/stu1907}, \href {https://ui.adsabs.harvard.edu/abs/2014MNRAS.445.2758R} {445, 2758}

\bibitem[\protect\citeauthoryear{{Rodrigues} et~al.,}{{Rodrigues} et~al.}{2017}]{2017MNRAS.467.1433R}
{Rodrigues} T.~S.,  et~al., 2017, \mn@doi [\mnras] {10.1093/mnras/stx120}, \href {https://ui.adsabs.harvard.edu/abs/2017MNRAS.467.1433R} {467, 1433}

\bibitem[\protect\citeauthoryear{{Skrutskie} et~al.,}{{Skrutskie} et~al.}{2010}]{2010SPIE.7735E..3HS}
{Skrutskie} M.~F.,  et~al., 2010, in {McLean} I.~S.,  {Ramsay} S.~K.,   {Takami} H.,  eds,  Society of Photo-Optical Instrumentation Engineers (SPIE) Conference Series Vol. 7735, Ground-based and Airborne Instrumentation for Astronomy III. p. 77353H, \mn@doi{10.1117/12.857724}

\bibitem[\protect\citeauthoryear{{Soummer}, {Pueyo}  \& {Larkin}}{{Soummer} et~al.}{2012}]{2012ApJ...755L..28S}
{Soummer} R.,  {Pueyo} L.,   {Larkin} J.,  2012, \mn@doi [\apjl] {10.1088/2041-8205/755/2/L28}, \href {https://ui.adsabs.harvard.edu/abs/2012ApJ...755L..28S} {755, L28}

\bibitem[\protect\citeauthoryear{{Tetzlaff}, {Neuh{\"a}user}  \& {Hohle}}{{Tetzlaff} et~al.}{2011}]{2011MNRAS.410..190T}
{Tetzlaff} N.,  {Neuh{\"a}user} R.,   {Hohle} M.~M.,  2011, \mn@doi [\mnras] {10.1111/j.1365-2966.2010.17434.x}, \href {https://ui.adsabs.harvard.edu/abs/2011MNRAS.410..190T} {410, 190}

\bibitem[\protect\citeauthoryear{{Vigan} et~al.,}{{Vigan} et~al.}{2021}]{2021A&A...651A..72V}
{Vigan} A.,  et~al., 2021, \mn@doi [\aap] {10.1051/0004-6361/202038107}, \href {https://ui.adsabs.harvard.edu/abs/2021A&A...651A..72V} {651, A72}

\bibitem[\protect\citeauthoryear{{da Silva} et~al.,}{{da Silva} et~al.}{2006}]{2006A&A...458..609D}
{da Silva} L.,  et~al., 2006, \mn@doi [\aap] {10.1051/0004-6361:20065105}, \href {https://ui.adsabs.harvard.edu/abs/2006A&A...458..609D} {458, 609}

\makeatother
\end{thebibliography}







\bsp	
\label{lastpage}
\end{document}